\documentclass[a4]{article}

\usepackage[lofdepth,lotdepth]{subfig}
\usepackage{graphicx}

\usepackage{algorithm}
\usepackage{algpseudocode}
\usepackage{color}
\usepackage{cite}
\usepackage[cmex10]{amsmath}
\usepackage{amssymb}
\usepackage{gensymb}
\usepackage{hyperref}
\usepackage{multicol}
\usepackage{enumitem}

\usepackage{color}
\usepackage{bm}
\usepackage[normalem]{ulem}

\newcommand{\bfm}[1]{\textbf{#1}}

\newcommand{\R}[1]{{\color{black}#1}}

\newcommand{\half}{\frac{1}{2}}
\newcommand{\dif}{{\rm d}}
\newcommand{\dvol}{{\rm d}^3\bfm{r}}
\newcommand{\dsur}{{\rm d}\bfm{s}}

\newcommand{\vJ}{\bfm{J}}

\newcommand{\vE}{\bfm{E}}
\newcommand{\vB}{\bfm{B}}
\newcommand{\vA}{\bfm{A}}

\newcommand{\vrh}{\hat{\bfm{r}}}
\newcommand{\vphh}{\hat{\bm{\varphi}}}
\newcommand{\Dt}{\Delta t}
\newcommand{\vDJ}{\Delta\bfm{J}}
\newcommand{\vDA}{\Delta\bfm{A}}

\newcommand{\vdelta}{\bm{\delta}}

\newcommand{\dtape}{d_{\rm tape}}
\newcommand{\vj}{\bfm{j}}
\newcommand{\Rsur}{R_{\rm sur}}
\newcommand{\Ohmmm}{$\Omega{\rm m}^2$}

\def\hyph{-\penalty0\hskip0pt\relax}

\begin{document}

\title{\bf Fast and accurate electromagnetic modeling of non-insulated and metal-insulated REBCO magnets\footnote{This is the Accepted Manuscript version of an article accepted for publication in Superconductor Science and Technology. IOP Publishing Ltd is not responsible for any errors or omissions in this version of the manuscript or any version derived from it.  The Version of Record is available online at DOI 10.1088/1361-6668/ad1c6f. This manuscript is hereby available under license CC BY-NC-ND.}}

\author{Enric~Pardo$^{1}$\footnote{Author to whom correspondence should be addressed (enric.pardo@savba.sk).}, Philippe~Fazilleau$^{2}$ \\
$^1$ Institute of Electrical Engineering, Slovak Academy of Sciences,\\Bratislava, Slovakia \\
$^2$ Universit\'e Paris-Saclay, CEA, 91191, Gif-sur-Yvette, France 
}

\maketitle

\begin{abstract}

REBCO high-temperature superconductors are promising for all\hyph superconducting high-field magnets, including ultra-high field magnets. Non-insulated (NI) and metal-insulated (MI) windings are a good solution for protection against electro-thermal quench. Design and optimization requires numerical modelling of REBCO inserts for high-field magnets. Here, we detail a fast and accurate two-dimensional (2D) cross-sectional model for the electromagnetic response of NI and MI coils, which is based on the Minimum Electro Magnetic Entropy Production (MEMEP). Benchmarking with an $A-V$ formulation method on a double pancake coil shows good agreement. We also analyse a fully superconducting 32 T magnet with a REBCO insert and a low-temperature superconducing (LTS) outsert. In particular, we analyze the current density, the screening curren induced field (SCIF), and the AC loss. We have shown that metal-insulated coils enable transfer of angular current in the radial direction, and hence magnet protection, while keeping the same screening currents and AC loss of \R{insulated} coils, even at relatively high ramp rates of 1 A/s. Surprisingly, soldered coils with low resistance between turns present relatively low AC loss for over-current configuration, which might enable higher generated magnetic fields. The numerical method presented here can be applied to optimize high-field magnets regarding SCIF in MI or NI magnets. It also serves as the basis for future electro-thermal modelling and multi-physics modeling that also includes mechanical properties.

\end{abstract}

\section{Introduction}

REBCO ($RE$Ba$_2$Cu$_3$O$_{7-x}$, where $RE$ is a rare earth like Y, Gd or Sm) high temperature (HTS) superconductors are very promising for high-field or ultra-high-field magnets, thanks to its high performance under high magnetic fields \cite{ugliettiD2019SST} and possibly higher operation temperatures than other superconductors \cite{mitchellN2021SST}. Static high-field magnets are required for a wide range of measurements of material research \cite{weijersHW2013IES, awajiS2017SST, hahnS2019Nat, huX2022SST, fazilleauP2020Cry, kimJ2020RSI, yanagisawaY2022SST} (including nuclear magnetic resonance (NMR) \cite{iwasaY2015IES, iguchiS2016SST, wikusP2022SSTa, yanagisawaY2022SST}), medical magnetic resonance imaging (MRI) \cite{vedrineP2015IES, parkinsonBJ2017SST}, particle accelerators and detectors \cite{vannugteren2018SST, wangX2019Ins}, and fusion \cite{mitchellN2021SST}, among other. 
	
For high-field magnets, it has been found that enabling current sharing between turns highly improves the reliability against thermal quench \cite{hahnS2012IES, yoonS2016SST, fazilleauP2018SST, lecrevisseT2022SSTa, mitchellN2021SST}. In addition, winding the coils without insulation (\R{no-insulation} (NI) coil \cite{hahnS2012IES, yoonS2016SST}) reduces the coil volume, and hence it enhances engineering current density and generated magnetic field. By using a moderately resistive metal tape between tapes, or metal-insulation (MI); magnet designers can adjust the turn-to-turn resistance to an optimum value \cite{fazilleauP2018SST, lecrevisseT2022SSTa}. Indeed, too low turn-to-turn resistance between tapes causes an undesired delay between the generated magnetic field and the input current \cite{yoonS2016SST}, which limits the magnet ramp speed. On the other hand, too high resistance between tapes may cause thermal stabilization issues. A stiff metal-insulation layer also improves the magnet mechanical properties. Another way to achieve optimized turn-to-turn resistance is to use the appropriate soldering material between turns \cite{mitchellN2021SST}. In this way, researchers can obtain turn-to-turn resistance from values comparable to NI down to orders of magnitude lower.

A key aspect in magnetic design are screening current effects, which degrade the magnetic field quality and stability \cite{iwasaY2015IES}, produce AC loss \cite{pardoE2016SST}, and increase hoop and other mechanical stresses \cite{xiaJ2019SST}, which could cause delamination. Although the effect of screening currents has been widely studied in \R{insulated} HTS windings, both by 2D cross-sectional models \cite{amemiyaN2008SST, yanagisawaY2010IES, pardoE2012SSTb, pardoE2013SST, pardoE2015SST, pardoE2016SST, xiaJ2015SST, xiaJ2019SST, berrospejuarezE2019SST, berrospejuarezE2020JCS, berrospejuarezE2021SST, wangY2020HiV} and 3D models \cite{uedaH2013IES, itohR2013PhC, uedaH2015IESa, mifuneT2019SST}, there are few studies for NI or MI magnets. A highly remarkable work about screening currents in HTS inserts is the 3D modeling of \cite{noguchiS2022SST}. Although 3D modeling can potentially describe all features of this geometry, such as low-critical-current section in a particular turn, 3D modeling is computationally expensive. The equivalent anisotropic homogeneous model in \cite{matairaR2020SST} for finite-element model (FEM) $H$-formulation in COMSOL software can reduce the 3D problem into its 2D cross-section. This 2D model enabled to describe screening currents in single NI pancake coils \cite{matairaR2020SST}, and stacks of 4 \cite{matairaR2021IES} and 8 \cite{zhongZ2022JSM} pancake coils with relatively low number of turns, around one order of magnitude lower than in HTS inserts in ultra-high field magnets. The reason seems to be still high computing times.

Magnet design requires fast and accurate modeling methods in order to make parameter sweeps and, potentially, serve as input for advanced optimization algorithms, such as those based on artificial intelligence \cite{yazdaniasramiM2023SST}. In this article, we develop a fast and accurate axi-symmetric numerical method for electro-magnetic modeling of MI, NI and soldered magnets, taking screening currents into account. We apply this numerical method to a 32 T full-superconducting magnet design, which is the baseline for further improved designs within the superEMFL project \cite{superEMFL}. In this model, we consider the magnetic field distribution generated by a 19 T LTS outsert magnet from Oxford Instruments, by taking the precise LTS winding cross-section into account. The computing time of the model is faster than the operation time-scales of the HTS insert, and hence it could be used for real-time operation. With this model, we systematically study the effect on screening currents, radial currents, generated field, screening current induced field (SCIF) and AC loss for a wide range of tape-to-tape \R{contact resistivity} between turns ($10^{-6}-10^{-10}$ \Ohmmm, or 10~000$-1~\mu\Omega{\rm cm}^2$), finding several interesting effects. In addition, we check the correctness of the model by a double-pancake benchmark with a completely different model, finding good agreement. \R{For notation simplicity, in this article we refer to ``AC loss" to any kind of dissipation, including Joule loss occurring above the critical current that also appears under DC currents.}

A substantial portion of this work was presented in the Magnet Technology conference in 2021 \cite{pardoE2021MT}, but no article has been published about this research.

\section{Studied configuration}

\begin{figure}[tbp]
\centering
{\includegraphics[trim=42 0 50 5,clip,width=10 cm]{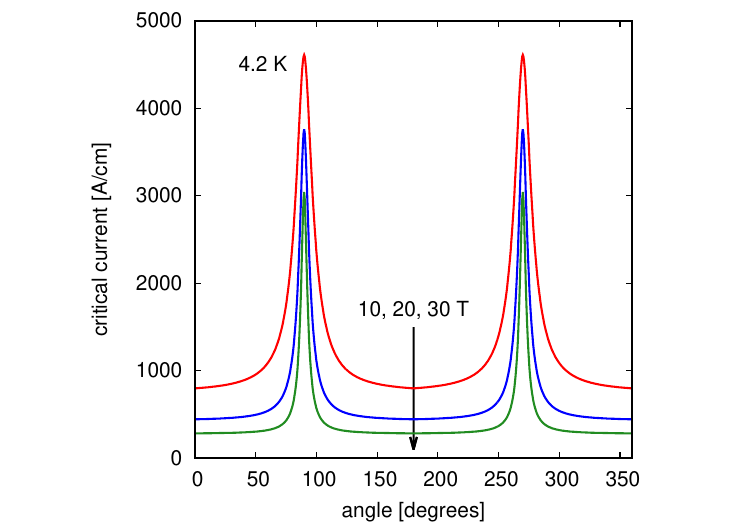}}
\caption{The model considers the \R{dependence of the critical current density on the magnetic flux density and its orientation,} $J_c(B,\theta)$, of Fujikura tape by means of the analytical fit of experimental data of \cite{fleiterJ2014CERN}. Above, angle $\theta=0$ corresponds to magnetic field perpendicular to the tape wide surface.}
\label{f.JcB}
\end{figure}

\begin{table}[tbp]
\caption{Main parameters of the REBCO HTS insert and the LTS outset. The calculated critical current of the REBCO insert is 371 A and 472 A for the Fujikura tape of \cite{fleiterJ2014CERN} and Theva APC tape, respectively. Thus, the current margin is 11 \% and 29 \%, respectively.}
\begin{tabular}{ll}
\hline 
\hline 
{\bf HTS insert} & \\
Generated magnetic field & 13 T \\
Rated current & 333 A \\
Inner diameter & 50 mm \\
Outer diameter & 102.5 mm \\
Tape width & 6 mm \\
Number of pancakes & 16 \\
Number of turns per pancake & 250 \\
Total thickness of copper stabilization & 20 $\mu$m \\
\hline
{\bf LTS insert} & \\
Generated magnetic field & 19 T \\
Cold bore diameter & 150 mm \\
Producer & Oxford Instruments \\
\hline
\hline 
\end{tabular}
\label{t.HTSLTS}
\end{table}

In this article, we study a tentative design for a 32 T magnet consisting on a REBCO high-temperature superconducting (HTS) insert proposed by CEA Saclay and a low-temperature superconducting (LTS) outsert provided by Oxford Instruments. The parameters of the HTS insert and LTS outsert are in table \ref{t.HTSLTS}. The HTS insert pre-design assumed the \R{dependence of the critical current density as a function of the magnetic flux density and its orientation,} $J_c(B,\theta)$, of Theva Advanced Pinning Center (APC) tape and 29 \% safety margin for the operation current (the insert critical current is 472 A), and maximum hoop stress below 600 MPa. However, in this article we assume the $J_c(B,\theta)$ dependence of Fujikura tape because of its complete $J_c(B,\theta,T)$ availability \cite{fleiterJ2014CERN} for future electro-thermal modelling\R{, where $T$ is the temperature}. For this tape, the insert critical current is 371 A (see section \ref{s.results} for details), and hence a safety margin of 11 \%.

We analyze the electro-magnetic properties and AC loss assuming several values of tape-to-tape \R{contact resistivity} in the radial direction: $\Rsur=10^{-6}$, $10^{-7}$, $10^{-8}$, $10^{-9}$, and $10^{-10}$~\Ohmmm~(or $R_{\rm sur}=$10~000, 1~000, 100, 10, 1 $\mu\Omega {\rm cm}^2$). The highest values ($10^{-6}$, $10^{-7}$ \Ohmmm) are typical of metal-insulated windings \cite{fazilleauP2018SST}, $10^{-8}$ \Ohmmm is of the order of magnitude of \R{no-insulation} pancake coils and $10^{-9}$, $10^{-10}$ \Ohmmm are relevant for soldered coils. In addition, we also consider electrically \R{insulated} windings, where the radial \R{contact resistivity} is infinite.

\section{Modeling Method}

In this section, we detail several aspects of the numerical methods used in this article. Sections \ref{s.axisym}-\ref{s.EJhom} refer to general considerations for both numerical methods and 3.4 (MEMEP) and 3.5 (A-V formulation) describe the numerical methods themselves. In this work, we use MEMEP for magnet calculations and both MEMEP and A-V formulation for benchmarking. 

For the MI REBCO insert of table \ref{t.HTSLTS}, the computing time for the evolution from \R{current} $I$=0 to the end of the initial ramp ($I$=333 A) is around 56 and 200 s for 10 and 20 elements per homogenized turn (25 homogenized turns per pancake) in a computer with a processor AMD Ryzen Threadripper PRO 3955WX 16-Cores. That is less than the real operation time of the magnet, being 333 s for a relatively high ramp of 1 A/s. Before the time evolution, the program required to compute the interaction matrices, which took additional 120 s for 20 elements per homogenized turn. Still, the total computing time of the MI coil falls below real operation.

\subsection{Effective axi-symmetrical model}
\label{s.axisym}

\begin{figure}[tbp]
\centering
{\includegraphics[trim=0 0 0 0,clip,width=10 cm]{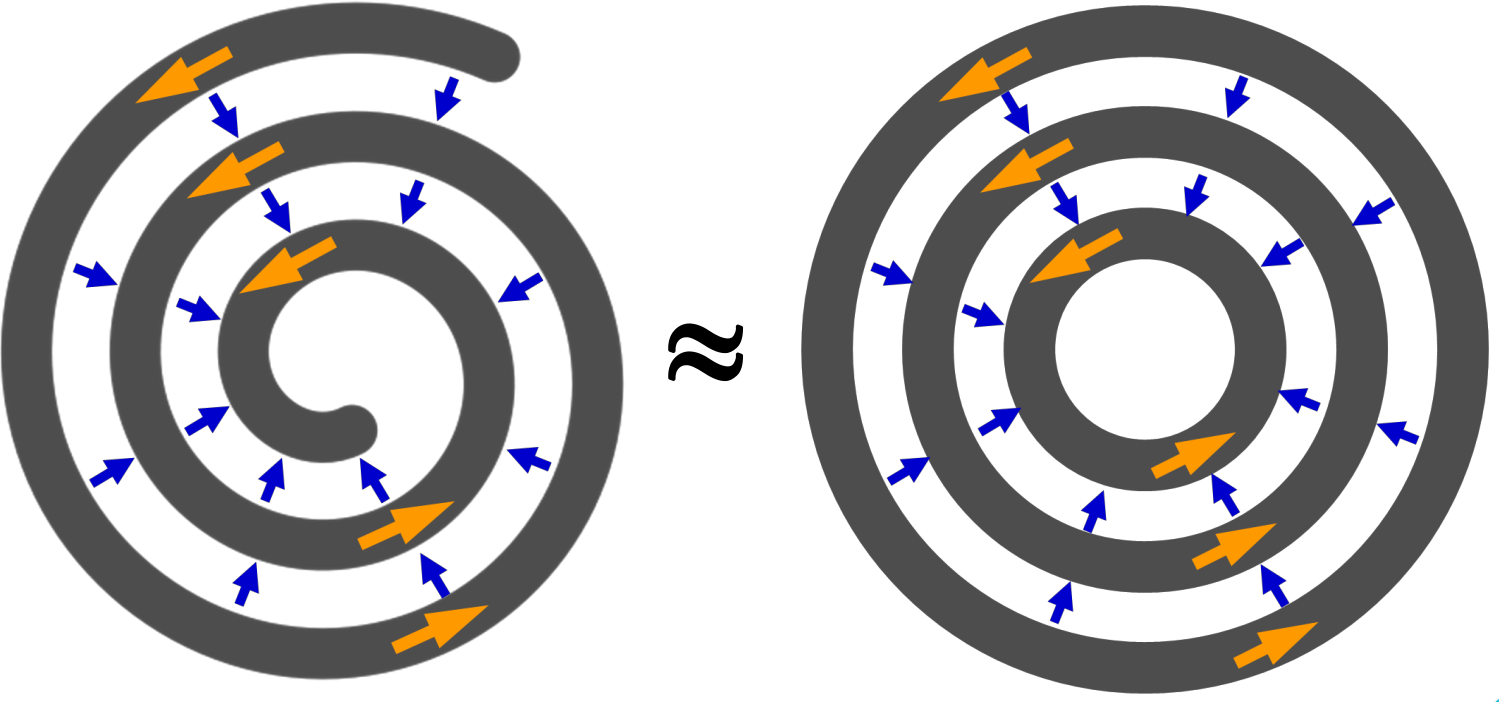}}
\caption{The spiral shape of no-insulation pancake coils of many turns can be well approximated by a set of concentric superconducting rings that are radially connected by normal conductor.}
\label{f.sketch}
\end{figure}

In this article, we assume an effective axi-symmetric model that enables radial currents due to finite \R{contact resistivity} between turns. The main assumptions is that each pancake coil has many turns and that the thickness of each turn is much smaller than the average radius. In this case, the radial current between neighboring turns is roughly independent on the angular coordinate, $\varphi$ (figure \ref{f.sketch}). Thus, both the radial and angular current densities are independent on the angular coordinate, $J_r(r,z)$ and $J_\varphi(r,z)$\R{; where $r$ and $z$ are the radial and axial coordinates}. Therefore, it is sufficient to model the 2D cross-section of each pancake coil. We ensure current conservation at each turn $k$ by imposing 
\begin{equation} \label{Icons}
I_{\varphi k}+I_{rk}=I,
\end{equation}
where $I$ is the coil input current and $I_\varphi$, $I_{rk}$ are the angular and radial currents at turn $k$.

We take all the inductive effects of the angular current into account, including screening currents. For simplicity, we neglect the direct inductive effects of the radial current density, and hence the magnetic field and vector potential that it creates. However, radial currents modify the angular currents through (\ref{Icons}), and hence radial currents have an indirect impact on inductive effects, such as the generated magnetic field. Indeed, there is a partial cancellation of the magnetic field and vector potential generated by the radial currents in each double pancake. The reason is that the direction of the radial current is opposite for each pancake of the double pancake. However, the plots of this article show the radial currents in all pancakes as having the same direction, for simplicity.

In order to numerically obtain $J_r$ and $J_\varphi$, we divide each turn cross-section into several rectangular elements. For simplicity, we assume that there is a single cell across each turn thickness, although we enable several cells across the turn (or tape) width. \R{We assume that both $J_\varphi$ and $rJ_r$ are uniform within the cross-section of each cell (with values $J_{\varphi i}$ and $r_iJ_{ri}$, where $r_i$ is the central radius of cell $i$), although we enable different values for each cell}. We impose uniform $rJ_r$ instead of $J_r$ in order to ensure that the radial current crossing any $r$-constant surface within the cell is the same.

\subsection{Electro-magnetic properties of \R{the superconducting layer}}
\label{s.EJs}

We assume that the relation between the current density and electric field in the \R{superconducting layer} follows a power-law $\vE(\vJ)$ relation \cite{grilliF2014IES, pardoE2023book}. In addition, the \R{superconducting layer} also presents a normal conducting component that can be considered in parallel to the purely superconducting carriers. Then, the superconducting current \R{density} follows $\vJ_s=\vJ_{ss}+\vJ_{sn}$, where $\vJ_{ss}$ is the contribution from purely superconducting carriers and $\vJ_{sn}$ is the normal contribution. Since both current densities are in parallel, the electric field in the superconductor obeys $\vE_s=\vE_{ss}+\vE_{sn}$, where $\vE_{ss}$ and $\vE_{sn}$ are the electric fields due to $\vJ_{ss}$ and $\vJ_{sn}$, respectively. These are
\begin{eqnarray}
&& \vE_{sn}=\rho_{sn}\vJ_{sn} \label{Esn} \\
&& \vE_{ss}=\frac{E_c}{J_c}\left ( \frac{\|\vJ_{ss}\|}{J_c} \right )^{n-1}\vJ_{ss}\equiv \rho_{ss}(\vJ_{ss})\vJ_{ss}, \label{Ess}
\end{eqnarray}
where $\rho_{sn}$ is the normal-carrier resistivity, $J_c$ is the critical current density, $n$ is the power-law exponent and $E_c$ is the critical current criterion, which we take as $E_c=10^{-4}$ V/m. In (\ref{Ess}), $\rho_{ss}$ is the superconducting-carrier resistivity. From (\ref{Esn}), (\ref{Ess}) and $\vJ_s=\vJ_{ss}+\vJ_{sn}$ the total current density follows
\begin{equation}
\vJ_s=\left [ \frac{1}{\rho_{ss}(\vJ_{ss})} + \frac{1}{\rho_{sn}} \right ] \vE_s.
\end{equation}
Next, we assume that for the evaluation of $\rho_{ss}$, $\vJ_{ss}\approx \vJ_s$. This approximation is valid for either $\|\vJ_{ss}\|\ll J_c$ because then $\rho_{ss}\ll \rho_{sn}$ and $\rho_{sn}$ can be ignored, or $\|\vJ_{ss}\|$ large enough to cause $\rho_{ss}\gg \rho_{sn}$ so that $\rho_{ss}$ can be ignored. Under this approximation, 
\begin{eqnarray}
\vE_s & \approx & \left [ \frac{1}{\rho_{ss}(\vJ_s)} + \frac{1}{\rho_{sn}} \right ]^{-1}\vJ_s
\nonumber \\
& = & \left [ \frac{J_c}{E_c} \left ( \frac{\|\vJ_{s}\|}{J_c} \right )^{1-n} + \frac{1}{\rho_{sn}} \right ]^{-1}\vJ_s
\equiv \rho_s(\vJ_s)\vJ_s , \label{EsJs}
\end{eqnarray}
where $\rho_s$ is the \R{overall} superconductor resistivity.

In general, $J_c$ and $n$ can depend on the magnetic field $\vB$; $J_c(\vB)$ and $n(\vB)$ or $J_c(B,\theta)$ and $n(B,\theta)$, where $B\equiv\|\vB\|$ and $\theta$ is the angle of $\vB$ with the normal to the superconductor surface. In this article, we use $J_c(B,\theta)$ of figure \ref{f.JcB}, $n=30$ and $\rho_{sn}$ approximated as that of stainless steel. \R{We choose this value because the normal state resistivity of the superconductor is of the same order of magnitude as stainless steel. Since the substrate is much thicker than the superconducting layer, the results will practically not depend on $\rho_{sn}$ when we also consider the substrate in the homogenized electro-magnetic properties of the whole tape (see section \ref{s.EJhom}). 
}

\subsection{Homogenized electro-magnetic properties}
\label{s.EJhom}

Next, we consider homogenized properties of the tapes and any metal-insulation or solder material between tapes. Although the models presented here could take all layers into account separately, homogenization highly enhances computing time. We consider that several layers of REBCO tapes are stacked in the radial direction, and hence the current in the radial direction is in series, while it is in parallel in the angular direction. We also take an additional \R{contact resistivity}, $\Rsur$, into account for the radial direction. Then, the homogenized engineering current density, $\vJ$, is related to the current in the superconductor, $\vJ_s$, and in each metal layer $i$, $\vJ_i$, as
\begin{eqnarray}
&& J_r=J_{sr}=J_{ir}=J_{{\rm sur},r} \quad \textrm{for all $i$} \label{Jr} \\
&& J_\varphi=J_{s\varphi}\frac{d_s}{\dtape} + \sum_{i=1}^{n_m}J_{i\varphi}\frac{d_i}{\dtape}, \label{Jph}
\end{eqnarray}
where $\dtape$ is the total tape thickness, $d_s$ is the superconducting layer thickness, $d_i$ is the thickness of metal layer $i$, and $n_m$ is the number of metal layers. Similarly, the homogenized electric field, $\vE$, follows
\begin{eqnarray}
&& E_r=E_{sr}\frac{d_s}{\dtape} + \sum_{i=1}^{n_m} E_{ir}\frac{d_i}{\dtape} + E_{{\rm sur},r} \label{Er}  \\
&& E_\varphi=E_{s\varphi}=E_{i\varphi} \quad \textrm{for all $i$}, \label{Eph}
\end{eqnarray}
where the contribution to the engineering electric field from the \R{contact resistivity} is $E_{{\rm sur},r}=\Rsur J_r/\dtape$. From (\ref{Jph}) and (\ref{Eph}) we find that
\begin{equation}
J_\varphi = \left [ \frac{1}{\rho_s(\vJ_s)}\frac{d_s}{\dtape} + \sum_{i=1}^{n_m}\frac{1}{\rho_i}\frac{d_i}{\dtape} \right ]^{-1} E_\varphi , 
\end{equation}
where $\rho_s(\vJ_s)$ is defined in (\ref{EsJs}). Similarly to section \ref{s.EJs}, for the evaluation of $\rho_s$ we assume that $J_{s\varphi}\approx J_\varphi({\dtape}/{d_s})$. Then, 
\begin{eqnarray}
E_\varphi & \approx & \left [ \frac{1}{\rho_s(J_r \vrh + J_\varphi({\dtape}/{d_s})\vphh) } \frac{d_s}{\dtape}  + \sum_{i=0}^{n_m} \frac{1}{\rho_i}\frac{d_i}{\dtape}  \right ]^{-1}J_\varphi \nonumber \\
& \equiv & \rho_\varphi(\vJ)J_\varphi, \label{EphJph}
\end{eqnarray}
where $\rho_\varphi(\vJ)$ is the homogenized resistivity in the angular direction. For the radial component, we obtain that
\begin{equation}
E_r=\left [ \rho_s(\vJ_s)\frac{d_s}{\dtape}  + \sum_{i=1}^{n_m} \rho_i\frac{d_i}{\dtape} + \frac{R_{\rm sur}}{d_{\rm tape}} \right ]J_r. \nonumber
\end{equation}
Again, we do the same approximation regarding $\rho_s$, resulting in 
\begin{eqnarray}
E_r & \approx & \left [ \rho_s(J_r\vrh + J_\varphi (\dtape/d_s)\vphh ) \frac{d_s}{\dtape} + \sum_{i=1}^{n_m} {\rho_i}\frac{d_i}{\dtape} + \frac{R_{\rm sur}}{d_{\rm tape}} \right ] J_r \nonumber \\
& \equiv & \rho_r(\vJ)J_r , \label{ErJr}
\end{eqnarray}
where $\rho_r$ is the homogenized resistivity in the radial direction.

\R{Up to now, we did not directly include the effect of the buffer layer, which is isolating. The turn-to-turn radial current will actually have to circumvent the buffer layer via the copper stabilization at the tape edges, reaching afterwards the superconductor and the copper layer on top of it. The increase of the turn-to-turn contact resistance due to this effect can be directly included in $R_{\rm sur}$ above. Actually, for the present configuration the contribution of $R_{\rm sur}$ to $\rho_r$ in (\ref{ErJr}) is orders of magnitude larger than the contribution of all the other metals, except for $R_{\rm sur}=10^{-10}$ \Ohmmm, where the metals contribute to around 40 \% of this value. In addition, the superconductor contribution, $\rho_s$, ranges from 0 to the normal-state resistivity, which is of the same order of magnitude as stainless steel. Then, in practice we could drop all the contributions of $\rho_r$ except of $R_{\rm sur}$. The detailed distribution of the radial current in the metals will not have an impact to the overall electromagnetic properties, as long as $R_{\rm sur}$ accounts for the experimental contact resistivity between turns.
}

Up to here, we homogenized the properties of a single tape, but we may further speed the computations up by homogenizing several neighboring turns together in a single element, as done in \cite{pardoE2013SST, zermeno2013JAP}. For this case, the homogenized $\vE(\vJ)$ relations of (\ref{EphJph}) and (\ref{ErJr}) are the same.

For the resistance of normal metals, we use the data provided in \cite{cryocomp}. The reader should note that for the stainless steel, $\rho$ depends on the temperature and for copper, $\rho$ depends on both temperature and the magnetic field. In this work we considered copper with RRR = 100.

\subsection{Variational method}
\label{s.MEMEP}

In this article, the calculations for the magnet and benchmark use the Minimum Electro-Magnetic Entropy Production (MEMEP) method \cite{pardoE2015SST, pardoE2016SST, pardoE2017JCP}. Following the three-dimensional (3D) formulation of \cite{pardoE2017JCP}, this method finds the current density $\vJ$ at time $t$ by minimizing the functional
\begin{eqnarray}
F[\vJ] & = & \int_\Omega \dvol \left \{  \frac{1}{2}\vDJ \cdot \frac{\vA[\vDJ]}{\Delta t} + \vDJ\cdot\frac{\vDA_a}{\Delta t} + U(\vJ) \right \} \nonumber \\
&& + \int_\Omega \dvol \nabla\phi \cdot \vJ, \label{FJ}
\end{eqnarray}
provided that the current density at a previous time $t-\Delta t$, $\vJ(t-\Delta t)$ is known, where $\Delta t$ is a certain time step. At the equation above \R{$\Omega$ is the region where the conductors and superconductors exist}, $\vDJ\equiv \vJ(t)-\vJ(t-\Dt)$, $\vA[\vDJ]$ is the vector potential in the Coulomb's gauge that $\vDJ$ generates \cite{grilliF2014IES, pardoE2023book}, $\vA_a$ is the vector potential in Coulomb's gauge created by external sources (or applied vector potential), $\phi$ is the electrostatic potential, and $U(\vJ)$ is the loss factor, defined as
\begin{equation} \label{U}
U(\vJ)\equiv \int_0^\vJ \dif\vJ'\cdot\vE(\vJ'),
\end{equation} 
which is valid for non-linear $\vE(\vJ)$ relations of the material, such as those in sections \ref{s.EJs} and \ref{s.EJhom}. The line integral of (\ref{U}) does not depend on the integration path for any physical $\vE(\vJ)$ relation, as shown in \cite{pardoE2017JCP}. In this work, we assume the initial condition of $\vJ=0$ at $t=0$.

Using that $\nabla\cdot(\phi\vJ)=\phi\nabla \cdot \vJ + \nabla\phi \cdot \vJ$ and that $\nabla\cdot\vJ=0$, the last term of (\ref{FJ}) becomes $\oint_{\partial\Omega}\dsur \cdot \phi\vJ$, where $\partial\Omega$ is the boundary of region $\Omega$ (its external surface) and $\dsur$ is its surface differential. If we assume that $\phi$ or $\vJ$ are uniform at the input and output leads of the magnet, this last term becomes
\begin{equation}
\int_\Omega \dvol \nabla\phi \cdot \vJ = \oint_{\partial\Omega}\dsur \cdot \phi\vJ = (\phi_f-\phi_i)I=-VI,
\end{equation}
where $I$ is the input current of the coil, $\phi_f$ and $\phi_i$ and the final and initial electrostatic potentials in the positive sense of the current and $V$ is the input voltage. In this paper, we take current constraints into account (the current $I$ is given). In addition, we assume that the current source is ideal, and hence the source provides any required $V$ in order to ensure the desired $I$. Thus, both $V$ and $I$ are fixed by the power source, and hence the $-VI$ term in the functional can be dropped if the only $\vJ(r,z)$ functions that we consider follow the current constraint, which is the case in this article.

For our axi-symmetrical approach (section \ref{s.axisym}), $\vJ$ does not depend on $\varphi$. In addition, we neglect the inductive effects of the radial current. Then, from (\ref{FJ}) we obtain that 
\begin{equation} \label{FJcyl}
F[\vJ]=2\pi\int_{\Omega_s}\dif r\dif z\ r \left \{ \half \Delta J_\varphi \frac{A_\varphi [\Delta J_\varphi]}{\Delta t} + \Delta J_\varphi \frac{\Delta A_{a\varphi}}{\Delta t} + U(\vJ)
\right \} , 
\end{equation}
where $\Omega_s$ is the cross-sectional region of the conducting or superconducting material. In this equation, we omitted the last term in (\ref{FJ}) because we consider input current constraint. We also used that in Coulomb's gauge $A_\varphi$ only depends on $J_\varphi$ \cite{grilliF2014IES, pardoE2023book}. 

Next, we discretize the problem by assuming that both $J_\varphi$ and $U(\vJ)$ are uniform in the cells. Then, 
\begin{equation} \label{FJi}
F\{\vJ_i\}= \sum_{i=1}^N 2\pi r_is_i\left \{ \half \Delta J_{\varphi i} 
\frac{\Delta A_{J\varphi i}}{\Delta t} + \Delta J_{\varphi i}\Delta A_{a\varphi i} + U(\vJ_i)
\right \} ,
\end{equation} 
where $\{\vJ_i\}$ are the values of $\vJ$ at all cells $i$, $r_i$ is the central radius of cell $i$, $s_i$ is its cross-section, $\vJ_i$ is the current density at the center of cell $i$ (there, $J_\varphi$ and $rJ_r$ are uniform with values $J_{\varphi i}$ and $r_iJ_{ri}$), $\Delta A_{J\varphi i}$ is the $\varphi$ component of the vector potential generated by $\Delta J_{\varphi}(r,z)$ averaged on the volume of cell $i$, which is
\begin{equation}
\Delta A_{J\varphi i} = \frac{1}{2\pi r_is_i}\int_{\Omega_i}\dvol A_\varphi[\Delta J_\varphi]
\end{equation}
\R{(where $\Omega_i$ is the region where cell $i$ exists),} and $A_{a\varphi i}$ is the same volume average but of $A_{a\varphi}$. Using Coulomb's gauge and axial symmetry, $A_{J\varphi i}$ is expressed as a function of $\{ J_{\varphi i}\}$ as
\begin{equation}
\Delta A_{J\varphi i} = \frac{1}{2\pi r_i}\sum_{j=1}^N J_{\varphi j} s_j C_{ij},
\end{equation}
where $N$ is the number of elements and $C_{ij}$ is the mutual inductance between loop $i$ and $j$, which are calculated as in \cite{pardoE2008SST}.

In the minimization process, we need to compute the change in $F$ due to a vector change $\vdelta$ in $\vJ$ at cell $i$, $\delta F_i(\vdelta)$. Using (\ref{FJi}), this quantity is
\begin{eqnarray}
\delta F_i(\vdelta) & \equiv & F\{ \vJ_j + \vdelta \delta_{ij} \} - F\{ \vJ_j \} \nonumber \\
& = & 2\pi r_is_i\delta_\varphi \frac{ \Delta A_{{\rm tot},\varphi i} }{\Delta t} + 
\frac{1}{2\Delta t} \delta_{\varphi}^2 s_i^2 C_{ii}+ 
2\pi r_i s_i\delta U_i(\vdelta) ,
\end{eqnarray}
where $\delta_{ij}$ is Kronecker delta, $A_{{\rm tot},\varphi i}$ is the $\varphi$ component of the total vector potential at cell $i$, $C_{ii}$ is the self-inductance of the current at the $\varphi$ direction at cell $i$, and 
\begin{eqnarray} \label{dU}
\delta U_i(\vdelta) & \equiv & U(\vJ_i+\vdelta) - U(\vJ_i) \nonumber \\
& = & \int_{\vJ_i}^{\vJ_i+\vdelta} \dif\vJ'\cdot \vE(\vJ') 
= \int_0^{\vdelta} \dif \bfm{j}\cdot \vE(\vJ_i+\bfm{j}).
\end{eqnarray}
A complication of $U(\vJ)$ for a general $\vE(\vJ)$ relation is that often there is no analytical integral for the $\delta U_i(\vdelta)$ expression of (\ref{dU}). This can be simplified if we consider small enough $\|\vdelta\|$ so that we can make a local linear approximation of $\vE(\vJ)$ for small changes in the current density, $\vj$, and hence
\begin{equation}
\vE(\vJ+\vj)\approx \vE(\vJ) + \frac{\vE(\vJ+\vdelta)-\vE(\vJ)}{\|\vdelta\|}\vj.
\end{equation}
Integrating (\ref{dU}) with this $\vE(\vJ)$ relation we obtain
\begin{equation}
\delta U_i(\vdelta)\approx \half\vdelta\cdot \left ( \vE(\vJ) + \vE (\vJ+\vdelta)
\right ).
\end{equation}

We minimize the functional in a similar way as in \cite{pardoE2015SST}, but with the difference that now we need to take both $J_\varphi$ and $J_r$ in the elements into account and comply with the current conservation equation of (\ref{Icons}). Simplifying, the minimization method follows the next steps.
\begin{enumerate}[label={\arabic*.}]
\item Increase the time by $\Delta t$ and input current by $\Delta I=I(t)-I(t-\Delta t)$. Then, add current density in the $\varphi$ direction in all turns that is uniformly distributed in each turn section so that the change in current corresponds to $\Delta I$.

\item Update $\Delta A_{Ji\varphi}$ for all cells $i$ caused by the change in current density in step 1. In the following steps, we modify $\vJ_i$ at all cells $i$ in order to take screening currents and radial currents into account.

\item Consider a small change in current $d>0$, which corresponds to the following changes in current density in the $\varphi$ and $r$ directions at any element $i$: $\delta_{i\varphi} = d/s_i$ and $\delta_{ir}=d/(2\pi r_i\Delta z_i)$, where $\Delta z_i$ is the size of element $i$ in the $z$ direction.

\item Calculate $\delta F_i(+\delta_{i\varphi}\vphh)$, $\delta F_i(-\delta_{i\varphi}\vphh)$, $\delta F_i(+\delta_{ir}\vrh)$, $\delta F_i(-\delta_{ir}\vrh)$ for all elements, $i$.

\item For each turn $k$, find the minimum $\delta F_i(+\delta_{i\varphi}\vphh)$ and the location of the minimum, $i_{k\varphi +}$ and $\delta F_{i_{k\varphi +}}$. Do the same for $\delta F_i(-\delta_{i\varphi}\vphh)$, $\delta F_i(+\delta_{ir}\vrh)$, $\delta F_i(-\delta_{ir}\vrh)$; obtaining $i_{k\varphi -}$ and $\delta F_{i_{k\varphi -}}$, $i_{kr+}$ and $\delta F_{i_{kr+}}$, $i_{kr+}$ and $\delta F_{i_{kr+}}$.

\item For all turns $k$, calculate
\begin{eqnarray*}
&& \delta F_{k,{\rm loop}}:= \delta F_{i_{k\varphi +}} + \delta F_{i_{k\varphi -}} \\
&& \delta F_{kr+}:= \delta F_{i_{kr+}} + \delta F_{i_{k\varphi -}} \\
&& \delta F_{kr-}:= \delta F_{i_{kr-}} + \delta F_{i_{k\varphi +}} .
\end{eqnarray*}
Each of the options above are consistent with (\ref{Icons}) because $I_{k\varphi}+I_{kr}$ remains unchanged.

\item For all turns $k$, identify the lowest of all 3 options above, $\delta F_{k,{\rm low}}$.

\item Identify the turn where $\delta_{k,{\rm low}}$ is the lowest and its value; $k_m$ and $\delta F_{k_m,{\rm low}}$, respectively.

\item {\bf If} $\delta F_{k_m,{\rm low}} < 0$ {\bf 	then}
\begin{enumerate}[label={\alph*.}]
\item Update cells in turn $k_m$, according to the lowest option in step 6.
\item Update $\Delta A_{J\varphi i}$ for all cells $i$ due to the change in current in step a.
\item Return to step 4. and repeat.
\end{enumerate}
{\bf else}
\begin{enumerate}[label={\alph*.}]
\item Return to step 1 and repeat until the desired final time is reached.
\end{enumerate}

\end{enumerate}

Actually, the minimization method is more refined in order to optimize it for speed and parallel computing, which uses sectors as in \cite{pardoE2016SST} and variable value of the current change, $d$, as in \cite{pardoE2015SST}. In addition, the magnetic field has an impact on $\vE(\vJ)$, since several parameters of this relation depend on $\vB$ (see sections \ref{s.EJs} and \ref{s.EJhom}). We take $J_c(\vB)$, $n(\vB)$ of the superconductor and $\rho(\vB)$ of all metals into account by solving $\Delta J$ and evaluating $\vB$ iteratively \cite{pardoE2015SST}. For the total magnetic field, we also compute the contribution from the LTS outsert. Given the cross-section shape and local engineering current density provided by Oxford Instruments, we numerically integrate the Biot-Savart law across the LTS for observation points at the center of each HTS element cross-section.

\subsection{A-V formulation}

Another way of computing screening currents in NI or MI windings is to use a modified version of the A-V formulation already published in \cite{dilasserG2017IES, fazilleauP2018IES, fazilleauP2022SST}. As shown in \cite{fazilleauP2022SST}, the $A-V$ formulation reduces to equation~\ref{AVform}.
\begin{equation}
\label{AVform}
   \dot{\vA}_J=-\vE(\vJ)-\dot{\vA}_a-\nabla \phi ,
\end{equation}
where the dot indicates partial time derivative and $\vA_J$, $\vA_a$ are the vector potential created by the current density and external sources, respectively. We modified the equivalent electrical circuit with the possibility for the current to radially bypass each element of the turn. Thermics can also been taken into account in order to calculate the Joule losses during transients or to study a quench. The computations have been performed with Matlab and the NAG libraries for integrating the stiff system of implicit ordinary differential equations.

\subsection{Benchmark}

\begin{figure}[tbp]
{\includegraphics[trim=0 0 0 0,clip,width=8 cm]{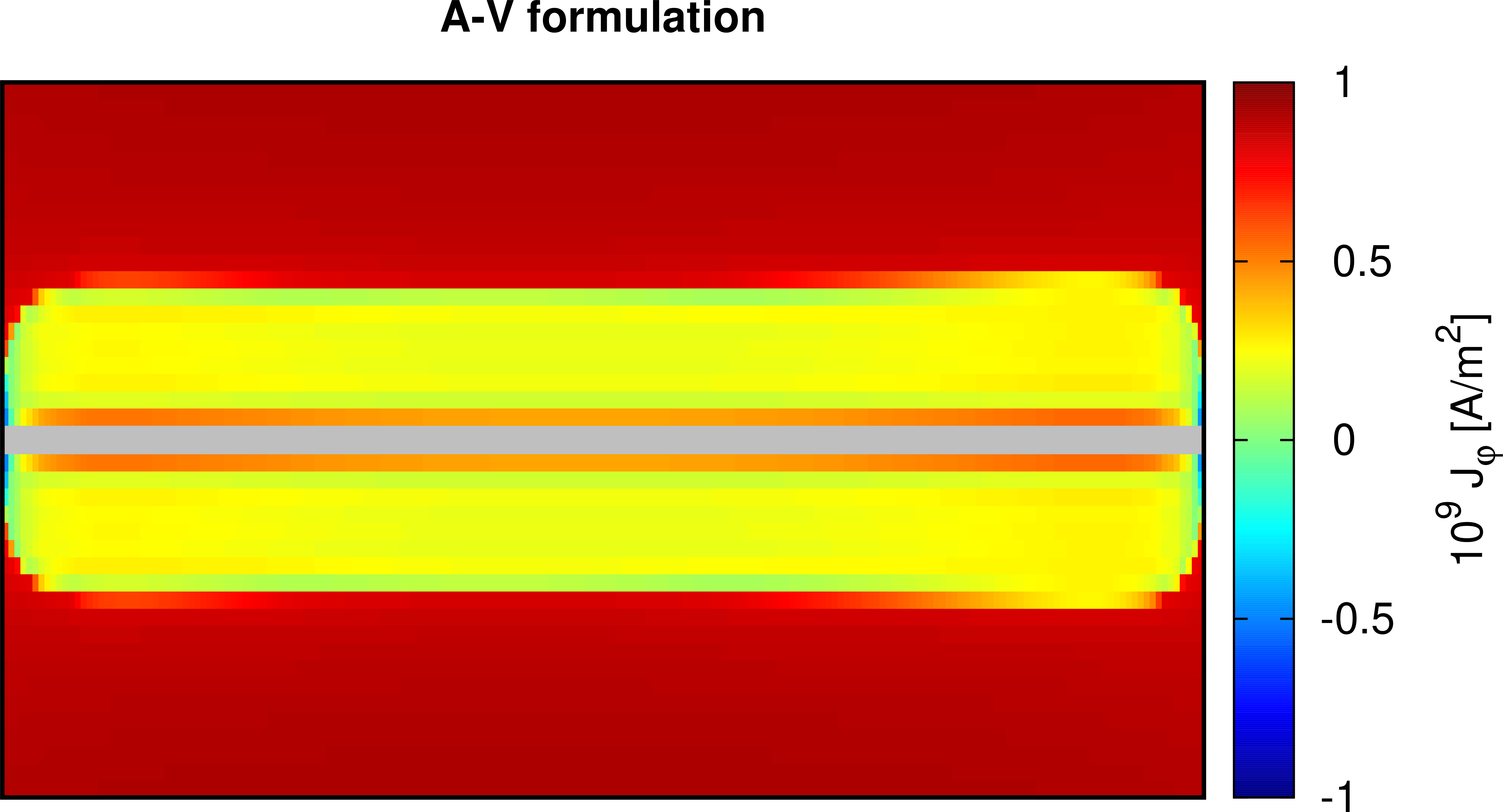}}
{\includegraphics[trim=0 0 0 0,clip,width=8 cm]{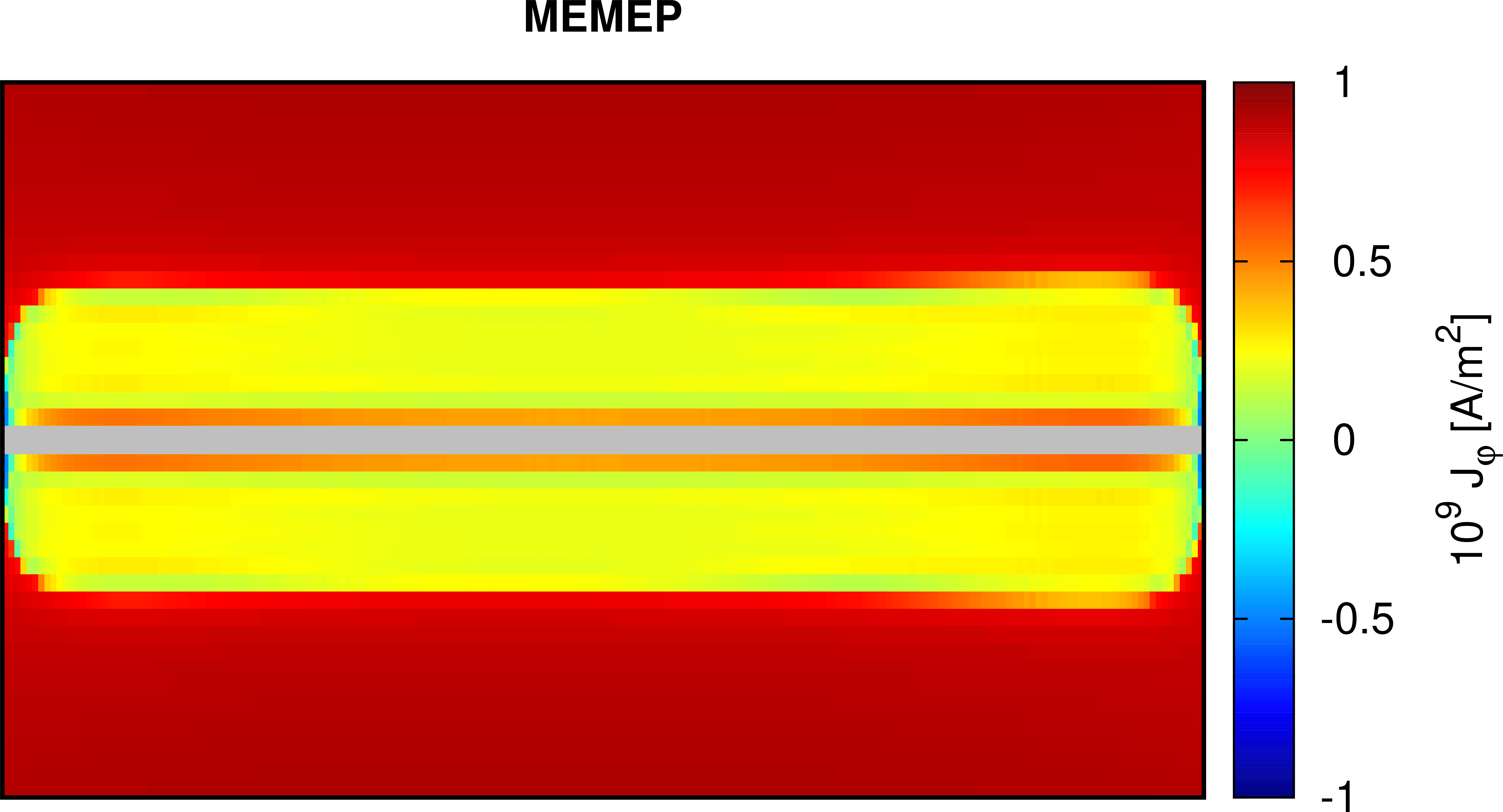}}\\
{\includegraphics[trim=0 0 0 0,clip,width=8 cm]{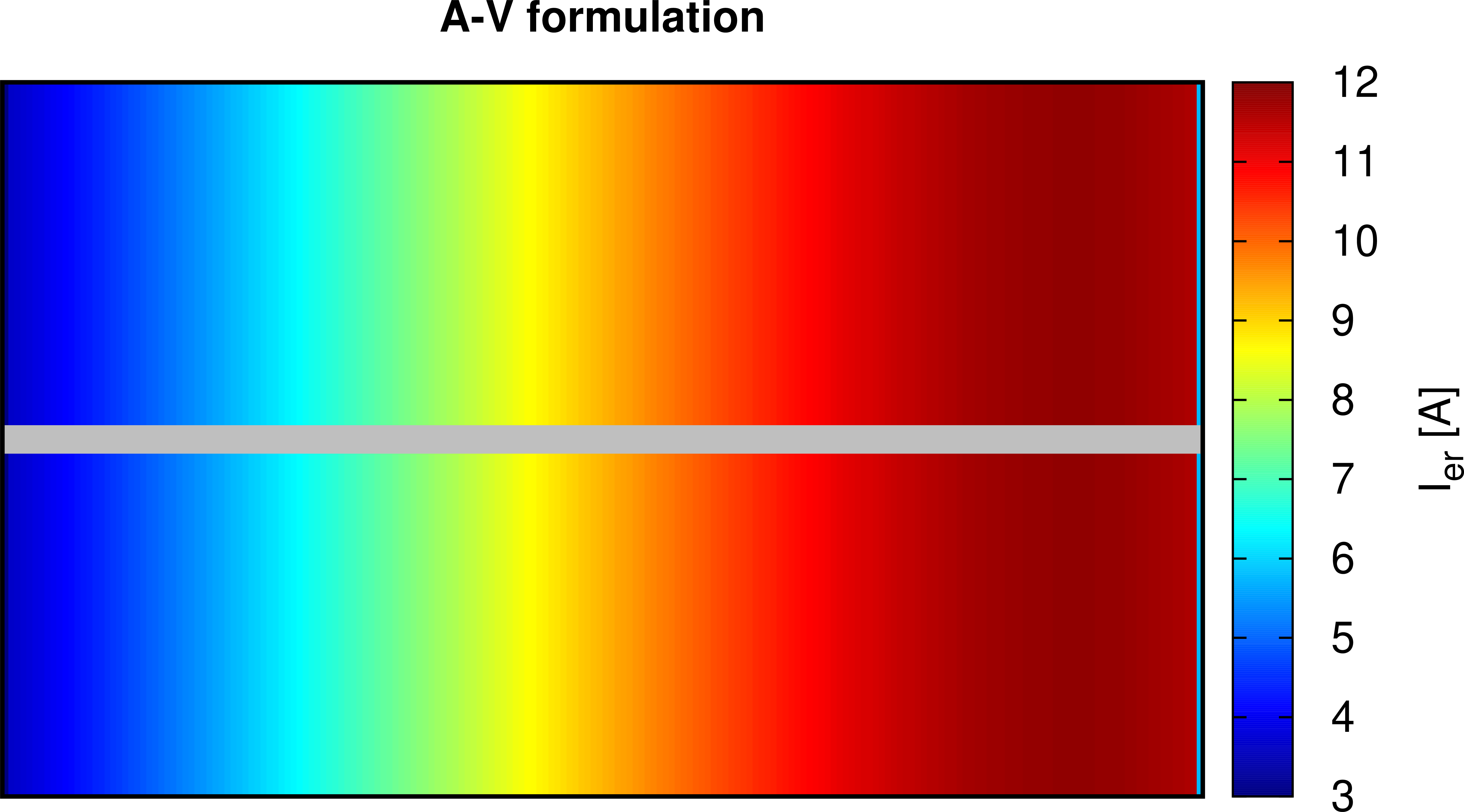}}
{\includegraphics[trim=0 0 0 0,clip,width=8 cm]{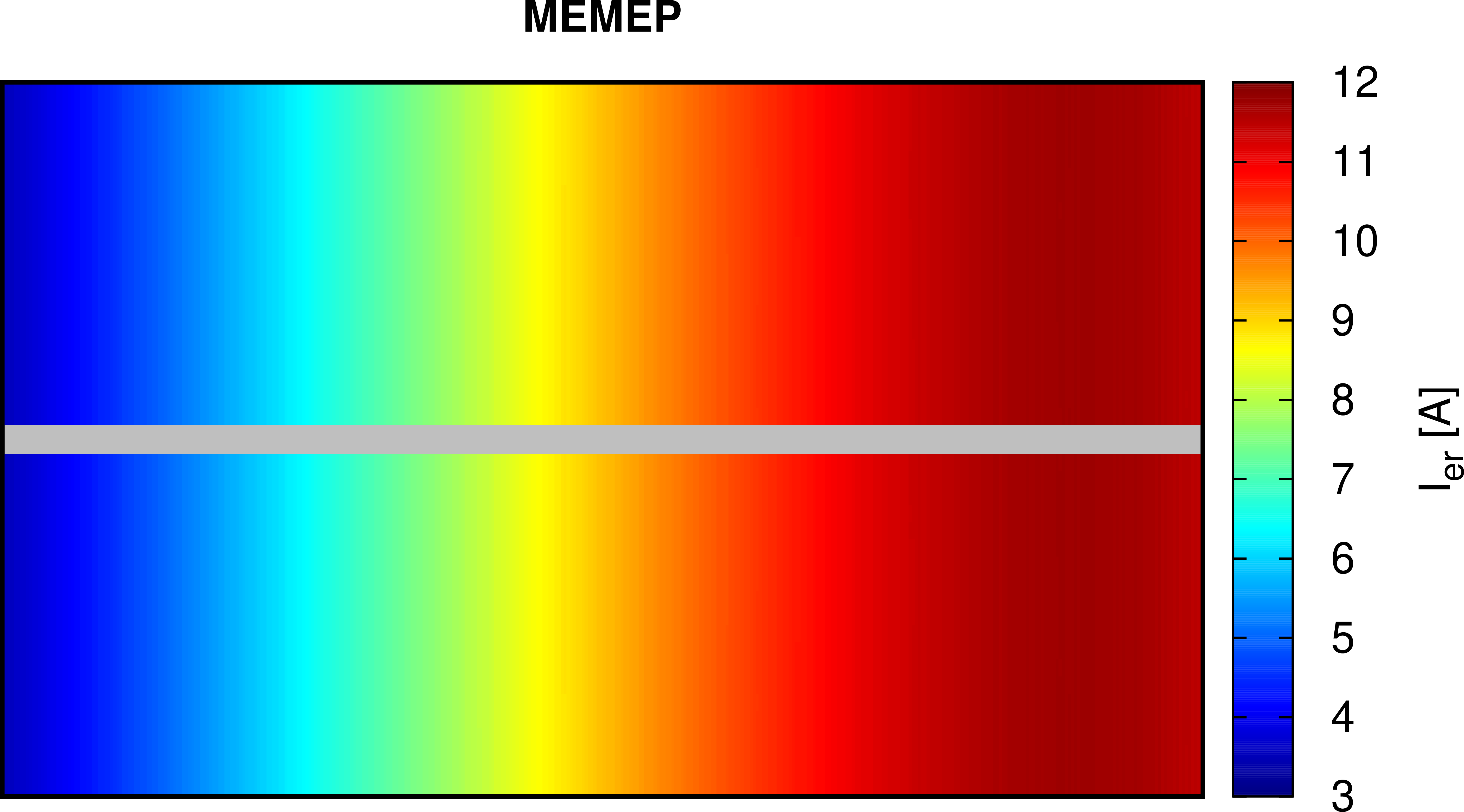}}
\caption{The computed homogenized angular current densities, $J_\varphi$, and turn-to-turn radial currents, $I_{er}$, by the two numerical methods agree, mutually validating both methods. The parameters of the benchmarked case are in table \ref{t.bench}.}
\label{f.Jbenchmark}
\end{figure}

\begin{table}[tbp]
\caption{Configuration studied for benchmarking the two numerical models. \R{We assume that the superconductor normal-state resistivity is the same as the substrate resistivity, for simplicity.}}
\begin{tabular}{ll}
\hline 
\hline 
{\bf Benchmark double pancake} & \\
Number of pancakes & 2 \\
Number of turns per pancake & 200 \\
Inner diameter & 50 mm \\
Separation between pancakes & 500 $\mu$m \\
Tape width & 6 mm \\
Tape and \R{insulation} thickness & 105 $\mu$m \\
Superconducting layer thickness & 1 $\mu$m \\
Critical current density & $10^{11}$ A/m$^2$ \\
Power-law exponent & 30 \\
Radial \R{contact resistivity} & $5\cdot 10^{-9}$ \Ohmmm \\
Substrate resistivity & 4.90$\cdot 10^{-7}$ $\Omega$m \\
Superconductor normal-state resistivity & 4.90$\cdot 10^{-7}$ $\Omega$m \\
Input current & 400 A \\
Ramp rate & 1 A/s \\
Number of elements per tape & 40 \\
\hline
\hline 
\end{tabular}
\label{t.bench}
\end{table}

In order to check the correctness of the models, we benchmark the MEMEP and $A-V$ formulation methods to each other on a double pancake coil, with parameters in table \ref{t.bench}. For the benchmark, we assume constant critical current density, $J_c$, power-law exponent, $n$, and resistivity (or normal-state resistivity for the superconductor) of all involved materials. Since we assume constant electric parameters, the background magnetic field is irrelevant for field cool of the REBCO winding. For each turn, we assume the homogenized $\vE(\vJ)$ relations of (\ref{EphJph}) and (\ref{ErJr}). \R{We represent $J_r$ as $I_{er}\equiv 2\pi rJ_rw_{\rm tape}$, where $w_{\rm tape}$ is the tape width in the $z$ direction. Thus, in case that $I_{er}$ is uniform in a certain turn, $I_{er}$ is the radial current at that turn.} As seen in figure \ref{f.Jbenchmark}, the results of $J_\varphi$ and $I_{er}$ for each model agree, which mutually validate both models. An interesting result is that $I_{er}$ in MEMEP is uniform at each turn, even though we allow non-uniform $I_{er}$ in the model. This is because we neglect direct inductive effects of $J_r$ (and $I_{er}$). The slightly lower current penetration at the right half (close to the outer radius) is due to higher radial current at that region.

\section{Results and discussion} \label{s.results}

\begin{figure}[tbp]
\centering
{\includegraphics[trim=0 0 0 0,clip,height=8 cm]{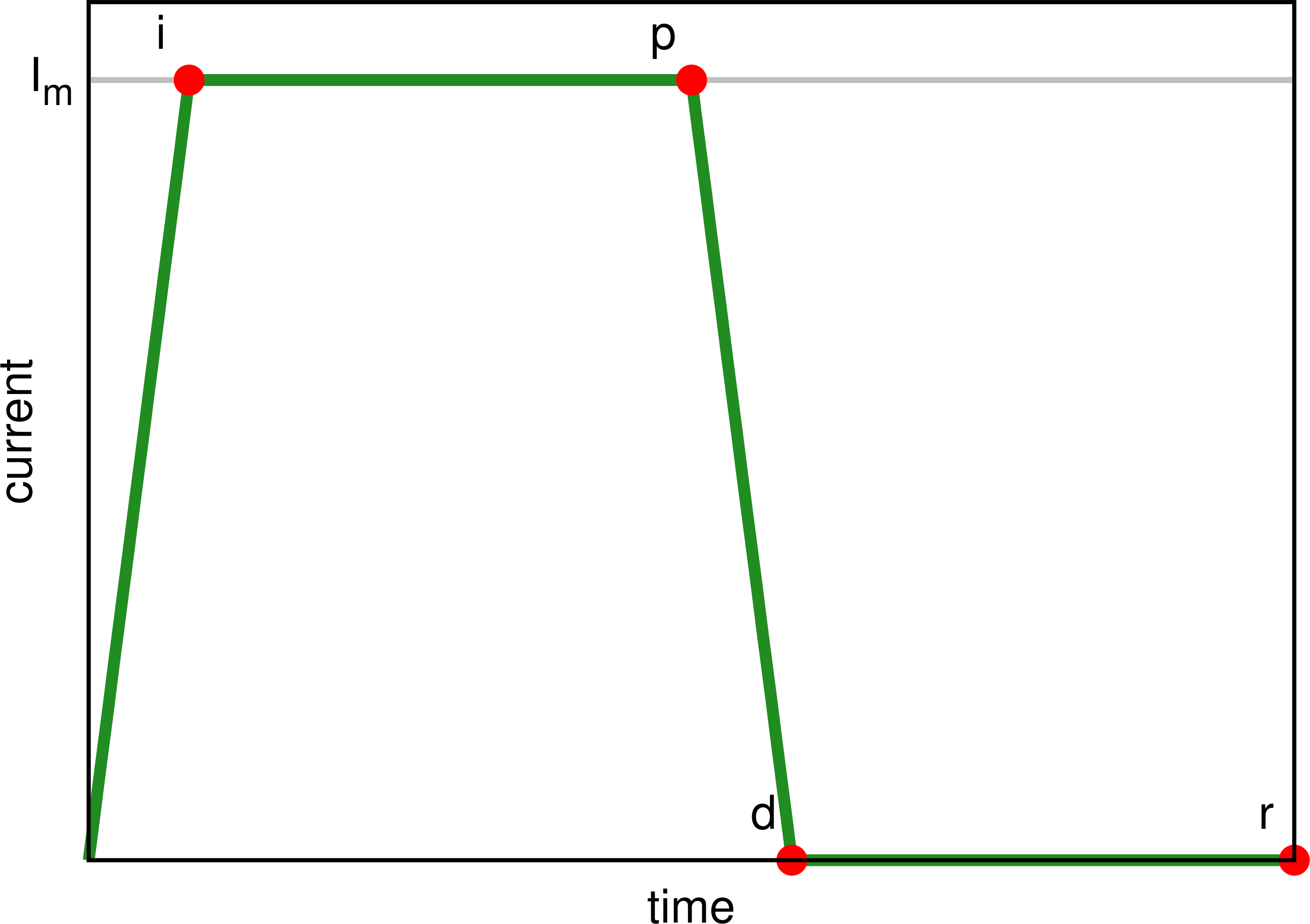}}
\caption{Sketch of the studied time dependence of the input current. Instants of special interest are the end of the increasing ramp, plateau, decreasing ramp, and relaxation (``i", ``p", ``d", ``r" above). The ramp increase or decrease is always of 1 A/s and the plateaus are of 3330 s.}
\label{f.It}
\end{figure}

Here, we present a detailed and systematic study of the electromagnetic behavior of the HTS insert of the 32 T magnet of in table \ref{t.HTSLTS} under the static magnetic field generated by the LTS outsert (table \ref{t.HTSLTS}). In particular, we consider the magnet operation profile of $I$ in the HTS insert of figure \ref{f.It}: ramp increase of 1 A/s until the desired maximum current, $I_m$, is obtained; a plateau of 3330 s; a ramp decrease of 1 A/s until $I=0$; and further relaxation of 3330 s. 

Before making the detailed analysis, we computed the insert critical current, which we define as the minimum current where $I$ reaches the critical current for at least one turn. We obtained this value by ramping up the magnet until the AC loss experiences a sharp rise. For this particular calculation, we used a power-law exponent of $n=1000$ and \R{insulated} turns. The obtained critical current is $I_c=371$~A. Below, we focus on three particular values of maximum current, $I_m=333$~A (operating current, 0.898 $I_c$), $I_m=466$~A (1.26 $I_c$), and $I_m=666$~A (1.80 $I_c$).

\subsection{Current density and radial currents}

Here, we analyze $\vJ(r,z)$ at the HTS insert at several time steps: increasing ramp, plateau, decreasing ramp, and relaxation (``i'', ``p'', ``d'', ``r'' in figure \ref{f.It}). 	Both $J_\varphi$ and $I_{er}$ are shown in figures \ref{f.Jph_ramp_333}-\ref{f.Jph_end_666}.

\subsubsection{Operational current ($I_m=333$~A).}

\begin{figure}[tbp]
{\includegraphics[trim=148 11 150 9,clip,height=8 cm]{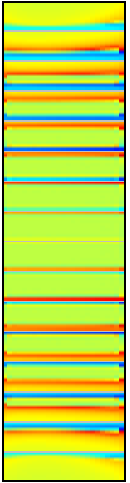}}
{\includegraphics[trim=148 11 150 9,clip,height=8 cm]{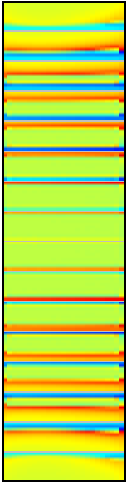}}
{\includegraphics[trim=148 11 150 9,clip,height=8 cm]{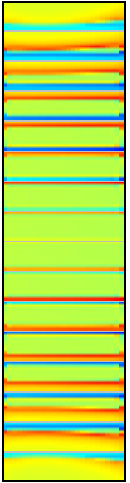}}
{\includegraphics[trim=148 11 94 9,clip,height=8 cm]{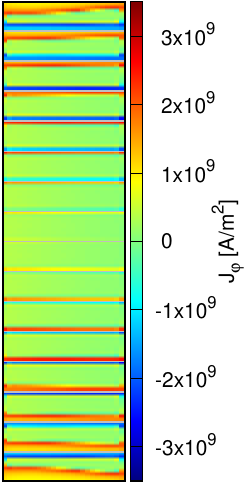}} \\
{\includegraphics[trim=148 8 150 9,clip,height=8.1 cm]{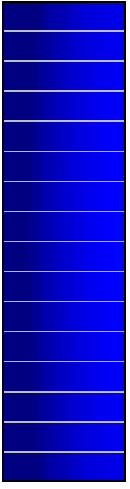}}
{\includegraphics[trim=148 8 150 9,clip,height=8.1 cm]{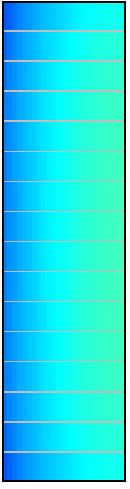}}
{\includegraphics[trim=148 8 150 9,clip,height=8.1 cm]{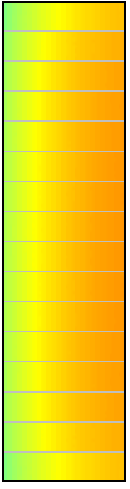}}
{\includegraphics[trim=148 8 102 9,clip,height=8.1 cm]{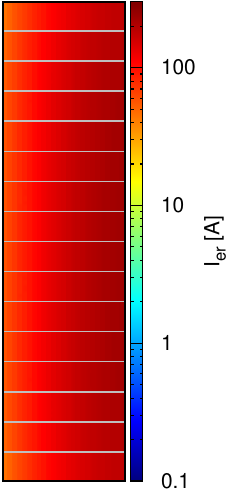}}
\caption{Angular current density, $J_\varphi$, and effective radial current, $I_{er}=2\pi rJ_rw_{\rm tape}$ at the end of the initial ramp (time ``i'' in figure \ref{f.It}) for $I_m=333$ A. The shown cross-sections are for turn-to-turn \R{contact resistivity} $\Rsur=10^{-6}$, $10^{-7}$, $10^{-8}$, $10^{-9}$ \Ohmmm from left to right. For each cross-section, the left and right edges are for the inner and outer radius, respectively.}
\label{f.Jph_ramp_333}
\end{figure}

The angular current density at the end of the initial ramp (figure \ref{f.Jph_ramp_333}) for $\Rsur= 10^{-6}$ and $10^{-7}$~$\Omega{\rm m}^2$ are practically the same, as well as for the insulated coil (not shown). Thus, for metal-insulated coils the screening currents are practically the same as for insulated coils. Thus, the screening currents for insulated coils are a good approximation for metal-insulated, which is relevant for computations that do not take radial currents into account. For non-insulated coils ($\Rsur=10^{-8}$~\Ohmmm) there is a slight decrease in $J_\varphi$, which becomes notorious for $\Rsur=10^{-9}$~\Ohmmm (soldered coil). The reason of this decrease is the transfer of angular current into radial current. Indeed, the radial current increases roughly inversely proportional to $\Rsur$ (see figure~\ref{f.Jph_ramp_333}), becoming comparable to the total current $I$ for $\Rsur=10^{-9}$~\Ohmmm. This transfer of current is due to inductive effects: the winding self-inductance requires a significant voltage to achieve the desired $I$; this creates a voltage between turns, and hence radial current. 

\begin{figure}[tbp]
\centering
{\includegraphics[trim=0 56 13 34,clip,width=14 cm]{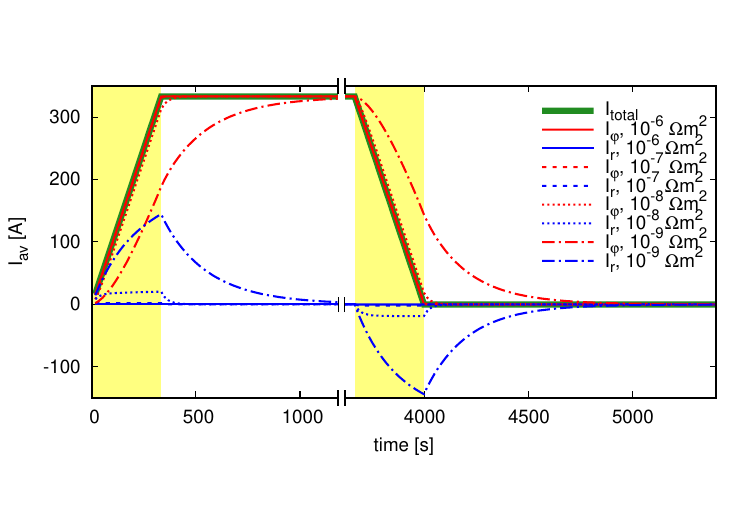}} \\
{\includegraphics[trim=0 56 13 34,clip,width=14 cm]{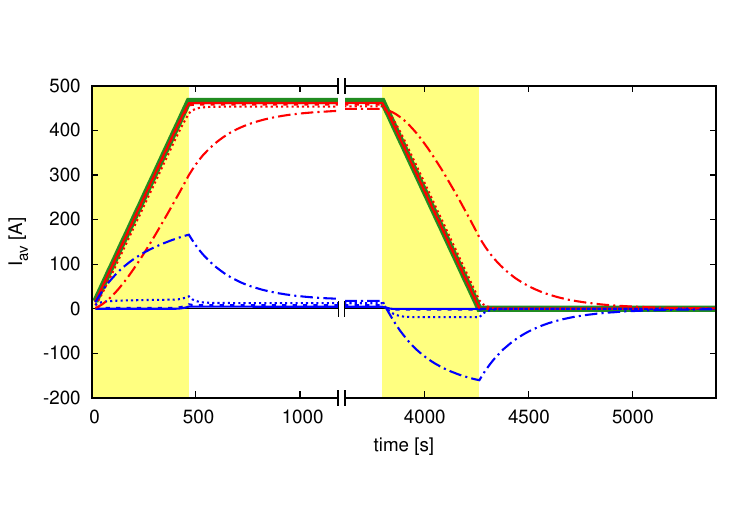}} \\
{\includegraphics[trim=0 30 13 34,clip,width=14 cm]{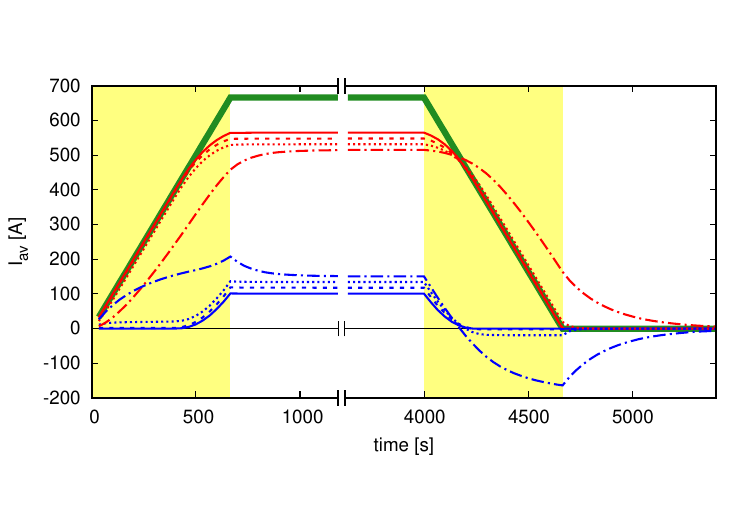}}
\caption{Average current in the REBCO insert cross-section in the angular and radial directions, $I_\varphi$ and $I_r$, respectively for several turn-to-turn \R{contact resistivity} (10$^{-6}$, 10$^{-7}$, 10$^{-8}$, 10$^{-9}$ $\Omega$m$^2$) and several maximum currents (333, 466, 666 A in graphs from top to bottom). Notice the broken horizontal scale. }
\label{f.Iav}
\end{figure}

The transfer of angular current into radial current becomes evident when looking at the average radial and angular current at the coil section:
\begin{eqnarray}
&& I_{\varphi,{\rm av}}=\frac{1}{n_t}\sum_{i=1}^N s_iJ_{\varphi i} \nonumber\\
&& I_{r,{\rm av}}=\frac{1}{n_t}\sum_{i=1}^N 2\pi\Delta z_i r_iJ_{ri} , 
\end{eqnarray}
where $n_t$ is the total number of turns and $\Delta z_i$ is the size of the cell $i$ in the $z$ direction. Indeed, for metal-insulated coils ($\Rsur=10^{-6}$ and $10^{-7}$~\Ohmmm), it follows that $I_{\varphi,{\rm av}}\approx I$ and $I_{r,{\rm av}}\approx 0$ (figure \ref{f.Iav}); for non-insulated coils ($\Rsur=10^{-8}$~\Ohmmm) there is a small but significant radial current, which causes a delay in the angular current; and for soldered coils this delay is so large that the radial current is almost as large as the angular one.

\begin{figure}[tbp]
{\includegraphics[trim=148 11 150 9,clip,height=8 cm]{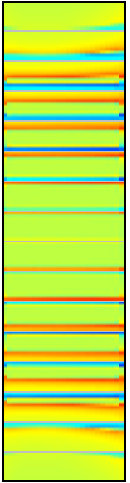}}
{\includegraphics[trim=148 11 150 9,clip,height=8 cm]{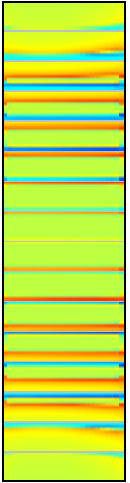}}
{\includegraphics[trim=148 11 150 9,clip,height=8 cm]{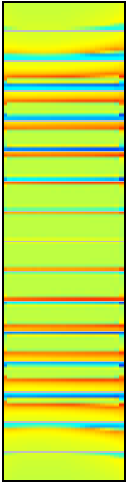}}
{\includegraphics[trim=148 11 94 9,clip,height=8 cm]{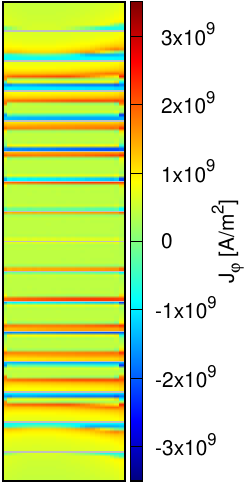}}
\caption{The same as figure \ref{f.Jph_ramp_333} but for the end of the plateau (time ``p'' in figure \ref{f.It}) and angular current density only ($I_m=333$ A and $\Rsur=10^{-6}$, $10^{-7}$, $10^{-8}$, $10^{-9}$ \Ohmmm from left to right).}
\label{f.Jph_rel_333}
\end{figure}

At the plateau (time ``p'' in figure \ref{f.It}), the radial current decreases exponentially until it vanishes at long enough times (figure \ref{f.Iav}). Then, $J_\varphi$ is almost the same for all the computed \R{contact resistivities}, and hence the screening currents always converge to those of the \R{insulated} case (see $J_{\varphi}$ at figure \ref{f.Jph_rel_333}).

\begin{figure}[tbp]
{\includegraphics[trim=148 11 150 9,clip,height=8 cm]{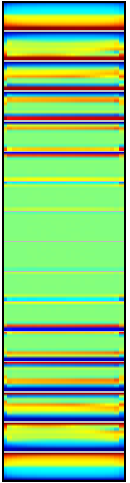}}
{\includegraphics[trim=148 11 150 9,clip,height=8 cm]{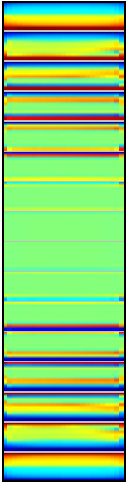}}
{\includegraphics[trim=148 11 150 9,clip,height=8 cm]{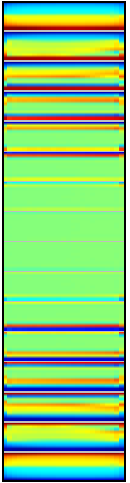}}
{\includegraphics[trim=148 11 94 9,clip,height=8 cm]{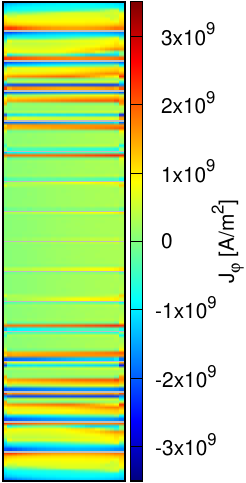}} \\
{\includegraphics[trim=148 8 150 9,clip,height=8.1 cm]{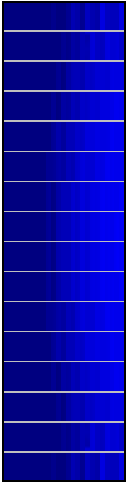}}
{\includegraphics[trim=148 8 150 9,clip,height=8.1 cm]{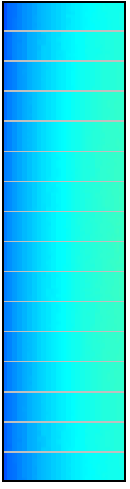}}
{\includegraphics[trim=148 8 150 9,clip,height=8.1 cm]{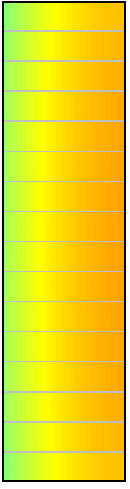}}
{\includegraphics[trim=148 8 102 9,clip,height=8.1 cm]{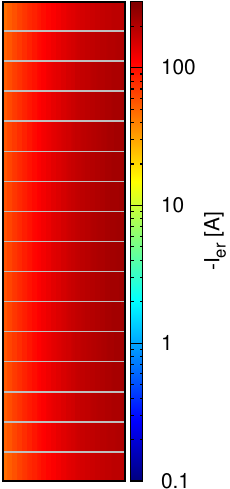}}
\caption{The same as figure \ref{f.Jph_ramp_333} but for the end of ramp down (time ``d'' in figure \ref{f.It}). Results for $I_m=333$ A and $\Rsur=10^{-6}$, $10^{-7}$, $10^{-8}$, $10^{-9}$ \Ohmmm from left to right.}
\label{f.Jph_down_333}
\end{figure}

The decreasing ramp presents a qualitatively similar behavior as the current transfer to the radial direction (not shown). Indeed, the screening currents at the end of the decreasing ramp (time ``d'' in figure \ref{f.It}) are very similar for the metal-insulated and non-insulated windings ($\Rsur=10^{-6}$, $10^{-7}$, $10^{-8}$~\Ohmmm) but are significantly lower for the soldered coil ($\Rsur=10^{-9}$~\Ohmmm) (see figure \ref{f.Jph_down_333}). Actually, at that time the radial currents are negative (figure \ref{f.Jph_down_333}), which causes $I_{\varphi,{\rm av}}>0$. Again, the radial currents are roughly proportional to the turn-to-turn \R{contact resistivity}.

\begin{figure}[tbp]
{\includegraphics[trim=148 11 150 9,clip,height=8 cm]{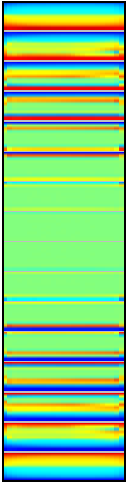}}
{\includegraphics[trim=148 11 150 9,clip,height=8 cm]{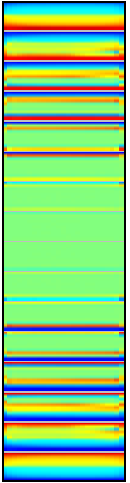}}
{\includegraphics[trim=148 11 150 9,clip,height=8 cm]{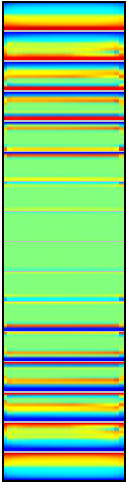}}
{\includegraphics[trim=148 11 94 9,clip,height=8 cm]{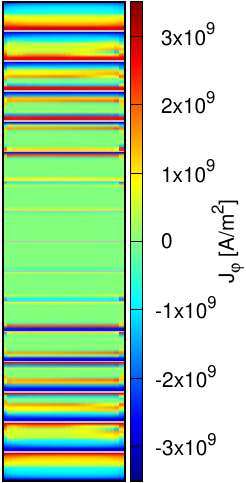}}
\caption{The same as figure \ref{f.Jph_ramp_333} but for the end relaxation (time ``r'' in figure \ref{f.It}) and angular current density only ($I_m=333$ A and $\Rsur=10^{-6}$, $10^{-7}$, $10^{-8}$, $10^{-9}$ \Ohmmm from left to right).}
\label{f.Jph_end_333}
\end{figure}

Finally, at the end of relaxation, the radial currents vanish and the screening currents are independent on the \R{contact resistivity} (see figure \ref{f.Jph_end_333}).

\subsubsection{Slightly above critical current ($I_m=1.26 I_c$).}

\begin{figure}[tbp]
{\includegraphics[trim=148 11 150 9,clip,height=8 cm]{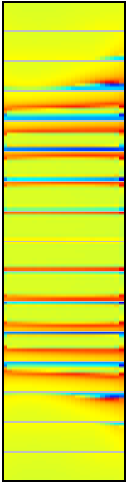}}
{\includegraphics[trim=148 11 150 9,clip,height=8 cm]{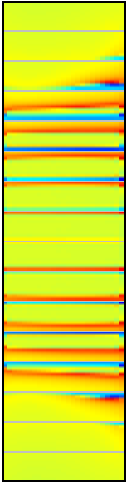}}
{\includegraphics[trim=148 11 150 9,clip,height=8 cm]{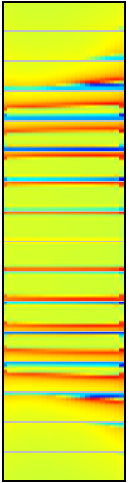}}
{\includegraphics[trim=148 11 94 9,clip,height=8 cm]{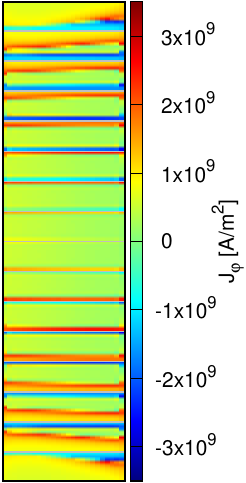}}\\
{\includegraphics[trim=148 8 150 9,clip,height=8.1 cm]{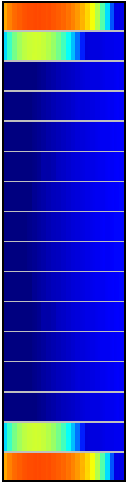}}
{\includegraphics[trim=148 8 150 9,clip,height=8.1 cm]{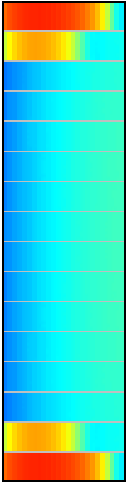}}
{\includegraphics[trim=148 8 150 9,clip,height=8.1 cm]{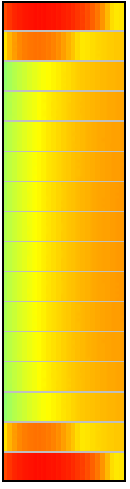}}
{\includegraphics[trim=148 8 102 9,clip,height=8.1 cm]{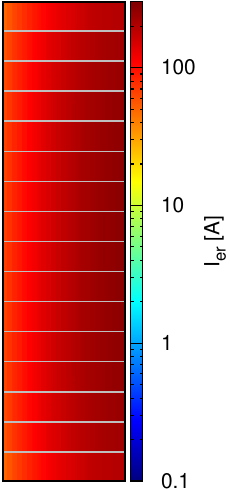}}
\caption{Angular current density, $J_\varphi$, and effective radial current, $I_{er}=2\pi rJ_rw_{\rm tape}$ at the end of the initial ramp (time ``i'' in figure \ref{f.It}) for $I_m=466$ A. The shown cross-sections are for $\Rsur=10^{-6}$, $10^{-7}$, $10^{-8}$, $10^{-9}$ \Ohmmm from left to right. For each cross-section, the left and right edges are for the inner and outer radius, respectively.}
\label{f.Jph_ramp_466}
\end{figure}

Next, we discuss the case of $I_m=466$~A, which is above the critical current (1.26 $I_c$).

\begin{figure}[tbp]
{\includegraphics[trim=148 11 150 9,clip,height=8 cm]{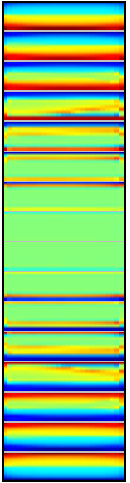}}
{\includegraphics[trim=148 11 150 9,clip,height=8 cm]{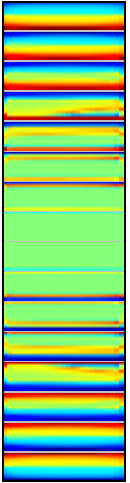}}
{\includegraphics[trim=148 11 150 9,clip,height=8 cm]{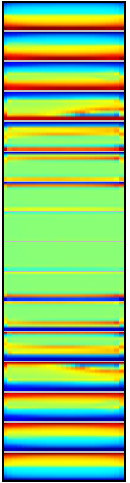}}
{\includegraphics[trim=148 11 94 9,clip,height=8 cm]{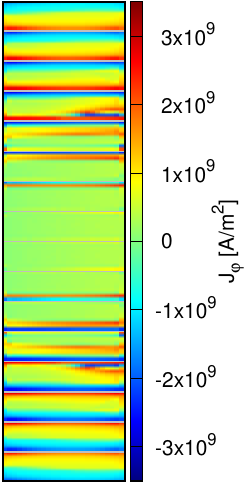}}\\
{\includegraphics[trim=148 8 150 9,clip,height=8.1 cm]{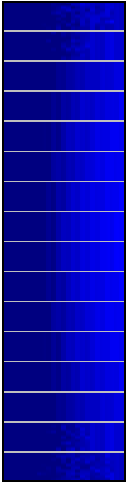}}
{\includegraphics[trim=148 8 150 9,clip,height=8.1 cm]{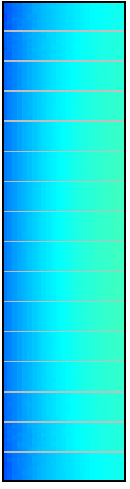}}
{\includegraphics[trim=148 8 150 9,clip,height=8.1 cm]{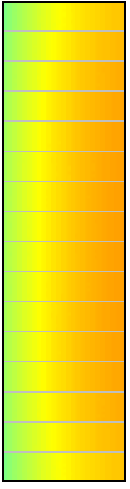}}
{\includegraphics[trim=148 8 102 9,clip,height=8.1 cm]{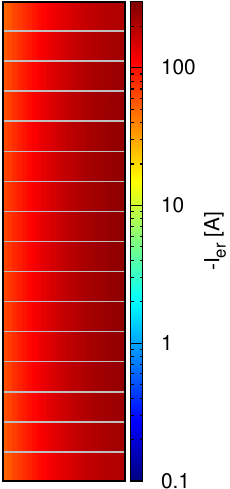}}
\caption{The same as figure \ref{f.Jph_ramp_466} but for the end of ramp down (time ``d'' in figure \ref{f.It}). Results for $I_m=466$ A and $\Rsur=10^{-6}$, $10^{-7}$, $10^{-8}$, $10^{-9}$ \Ohmmm from left to right.}
\label{f.Jph_down_466}
\end{figure}

At the end of the initial ramp (time ``i'' in figure \ref{f.It}), $J_\varphi$ is qualitatively similar to the case of $I_m=466$~A (compare figure \ref{f.Jph_ramp_333} and \ref{f.Jph_ramp_466}). The main difference is that $J_\varphi$ is positive at the whole section of most of the turns at the top and bottom double pancakes for the metal-insulated and non-insulated coils ($\Rsur=10^{-6}$, $10^{-7}$, $10^{-8}$~\Ohmmm) (see figure \ref{f.Jph_ramp_466}). This confirms that the superconductor is saturated or with $I$ above $I_c$ in some places. The behavior of $I_{er}$ for $I_m=466$~A is very different from the sub-critical configuration of $I_m=333$~A. Indeed, there appear significant radial currents at the top and bottom double pancakes for both metal-insulated and non-insulated coils ($\Rsur=10^{-6}$, $10^{-7}$, $10^{-8}$~\Ohmmm). This is because $J_\varphi$ becomes over-critical at most turns of these pancakes. Then, the superconductor resistivity dramatically increases, causing a transfer of current in the radial direction. Consistently, there is a slight increase in $I_{er}$ at the top and bottom double pancakes with decreasing $\Rsur$ from $10^{-6}$ to $10^{-8}$~\Ohmmm. The soldered coil ($\Rsur=10^{-9}$~\Ohmmm) presents roughly the same $I_{er}$ for all pancakes, which is typical of inductive behavior like for $I_m=333$~A (see figures \ref{f.Jph_ramp_333} and \ref{f.Jph_down_466}).

\begin{figure}[tbp]
{\includegraphics[trim=148 11 150 9,clip,height=8 cm]{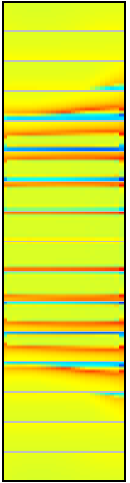}}
{\includegraphics[trim=148 11 150 9,clip,height=8 cm]{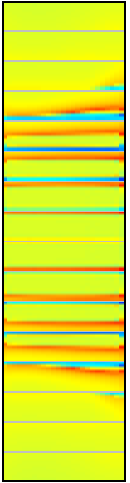}}
{\includegraphics[trim=148 11 150 9,clip,height=8 cm]{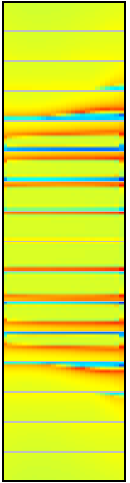}}
{\includegraphics[trim=148 11 94 9,clip,height=8 cm]{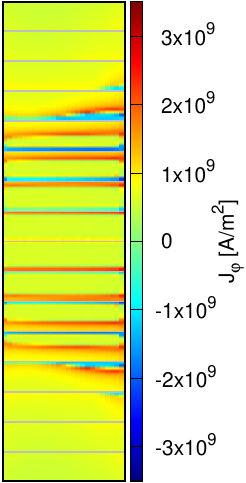}}\\
{\includegraphics[trim=148 8 150 9,clip,height=8.1 cm]{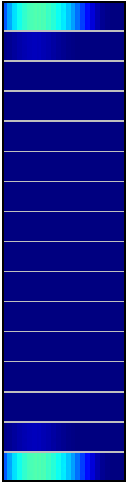}}
{\includegraphics[trim=148 8 150 9,clip,height=8.1 cm]{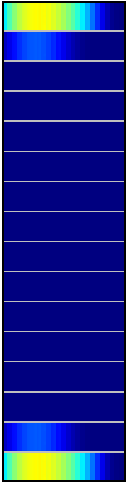}}
{\includegraphics[trim=148 8 150 9,clip,height=8.1 cm]{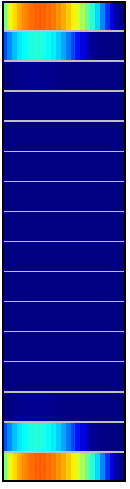}}
{\includegraphics[trim=148 8 102 9,clip,height=8.1 cm]{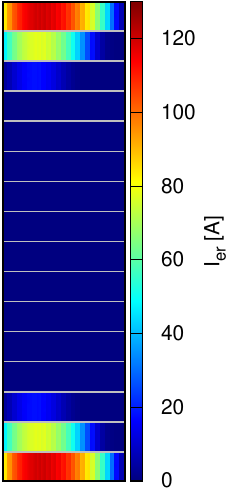}}
\caption{The same as figure \ref{f.Jph_ramp_466} but for the end of the plateau (time ``p'' in figure \ref{f.It}). Results for $I_m=466$ A and $\Rsur=10^{-6}$, $10^{-7}$, $10^{-8}$, $10^{-9}$ \Ohmmm from left to right.}
\label{f.Jph_rel_466}
\end{figure}

At the end of the plateau (time ``p'' of figure \ref{f.It}), $J_\varphi(r,z)$ is almost the same for all \R{contact resistivities}. However, the penetration of screening currents at the 4th and 5th pancakes from top and bottom slightly increases with decreasing $\Rsur$ (figure \ref{f.Jph_rel_466}). The reason is that $I_{er}$ at the top and bottom double pancakes increases with decreasing $\Rsur$ (figure \ref{f.Jph_rel_466}), causing a decrease in $I_\varphi$ at the turns of these pancakes. This increases the radial magnetic field at the inner pancakes, increasing the penetration of screening currents.

The essence of the time evolution of the angular and radial currents can be seen in figure \ref{f.Iav}. The main features are that: now, there is radial current at the plateau for all \R{contact resistivities}; the average radial current approaches to a constant value at the end of the plateau and this value decreases with $\Rsur$.

At the end of the current ramp down, there appears negative $I_{er}$ for all $\Rsur$ that is roughly uniform (figure \ref{f.Jph_down_466}), which is caused by inductive effects. This $I_{er}$ decreases the angular currents for low $\Rsur$, such as $10^{-9}$~\Ohmmm (figure \ref{f.Jph_down_466}). 

\begin{figure}[tbp]
{\includegraphics[trim=148 11 150 9,clip,height=8 cm]{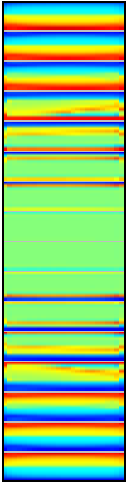}}
{\includegraphics[trim=148 11 150 9,clip,height=8 cm]{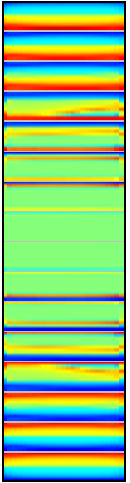}}
{\includegraphics[trim=148 11 150 9,clip,height=8 cm]{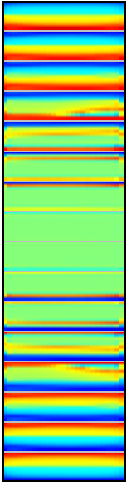}}
{\includegraphics[trim=148 11 94 9,clip,height=8 cm]{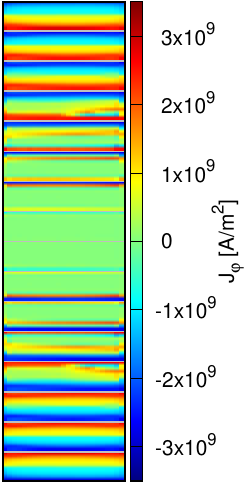}}
\caption{The same as figure \ref{f.Jph_ramp_466} but for the end relaxation (time ``r'' in figure \ref{f.It}) and angular current density only ($I_m=466$ A and $\Rsur=10^{-6}$, $10^{-7}$, $10^{-8}$, $10^{-9}$ \Ohmmm from left to right).}
\label{f.Jph_end_466}
\end{figure}

At the end of relaxation, the screening currents are almost the same for all $\Rsur$, except for small differences at inner pancakes because of differences in $J_\varphi$ at the end for the plateau (figure \ref{f.Jph_end_466}).

\subsubsection{Well above critical current ($I_m=1.80 I_c)$.}

\begin{figure}[tbp]
{\includegraphics[trim=148 11 150 9,clip,height=8 cm]{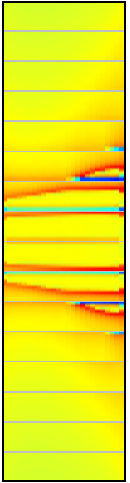}}
{\includegraphics[trim=148 11 150 9,clip,height=8 cm]{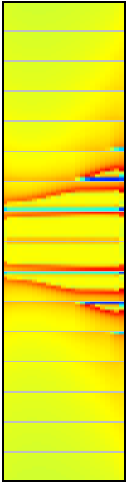}}
{\includegraphics[trim=148 11 150 9,clip,height=8 cm]{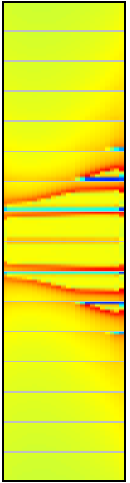}}
{\includegraphics[trim=148 11 94 9,clip,height=8 cm]{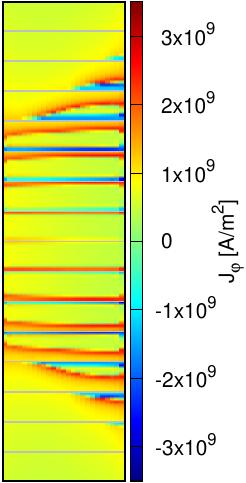}} \\
{\includegraphics[trim=148 8 150 6,clip,height=8.2 cm]{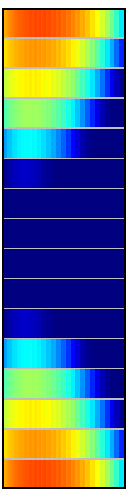}}
{\includegraphics[trim=148 8 150 6,clip,height=8.2 cm]{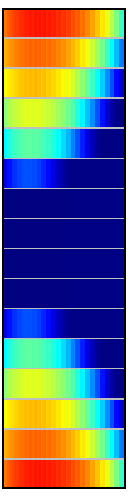}}
{\includegraphics[trim=148 8 150 6,clip,height=8.2 cm]{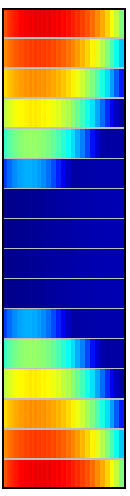}}
{\includegraphics[trim=148 8 102 6,clip,height=8.2 cm]{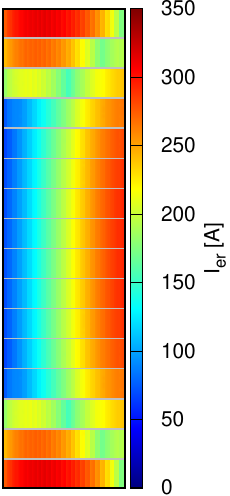}}
\caption{Angular current density, $J_\varphi$, and effective radial current, $I_{er}=2\pi rJ_rw_{\rm tape}$ at the end of the initial ramp (time ``i'' in figure \ref{f.It}) for $I_m=666$ A. The shown cross-sections are for $\Rsur=10^{-6}$, $10^{-7}$, $10^{-8}$, $10^{-9}$ \Ohmmm from left to right. For each cross-section, the left and right edges are for the inner and outer radius, respectively.}
\label{f.Jph_ramp_666}
\end{figure}

Here, we analyze $\vJ(r,z)$ for $I_m=666$~A, which is well above the critical current ($I_m=1.80I_c$).

At the end of the increasing ramp (time ``i'' in figure \ref{f.It}), more than half of the coil section does not present any screening current (figure \ref{f.Jph_ramp_666}), which suggests that $I$ exceeds $I_c$ of most of the turns. Again, the case of $\Rsur=10^{-9}$~\Ohmmm presents lower saturation because of the delay in the angular currents due to the large inductive radial currents (figure \ref{f.Jph_ramp_666}). For the metal-insulated and non-insulated windings ($\Rsur=10^{-6}$, $10^{-7}$, $10^{-8}$~\Ohmmm), $I_{er}$ is significant for all pancakes except for the four central ones. 

\begin{figure}[tbp]
{\includegraphics[trim=148 11 150 9,clip,height=8 cm]{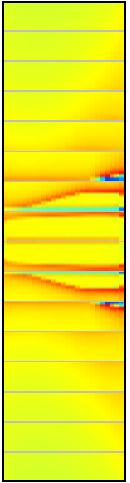}}
{\includegraphics[trim=148 11 150 9,clip,height=8 cm]{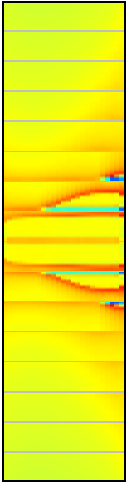}}
{\includegraphics[trim=148 11 150 9,clip,height=8 cm]{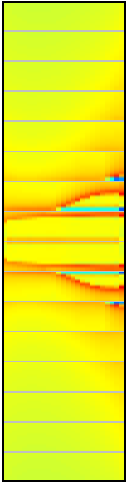}}
{\includegraphics[trim=148 11 94 9,clip,height=8 cm]{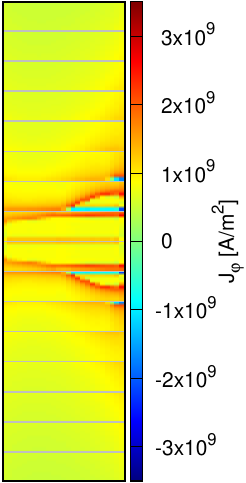}}
\caption{The same as figure \ref{f.Jph_ramp_666} but for the end of the plateau (time ``p'' in figure \ref{f.It}) and angular current density only ($I_m=666$ A and $\Rsur=10^{-6}$, $10^{-7}$, $10^{-8}$, $10^{-9}$ \Ohmmm from left to right).}
\label{f.Jph_rel_666}
\end{figure}

\begin{figure}[tbp]
{\includegraphics[trim=148 11 150 9,clip,height=8 cm]{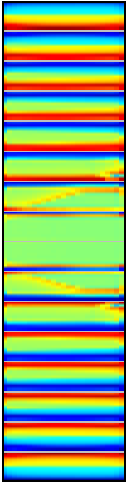}}
{\includegraphics[trim=148 11 150 9,clip,height=8 cm]{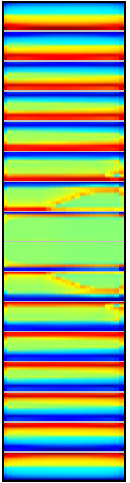}}
{\includegraphics[trim=148 11 150 9,clip,height=8 cm]{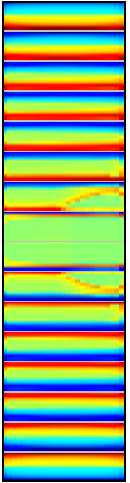}}
{\includegraphics[trim=148 11 94 9,clip,height=8 cm]{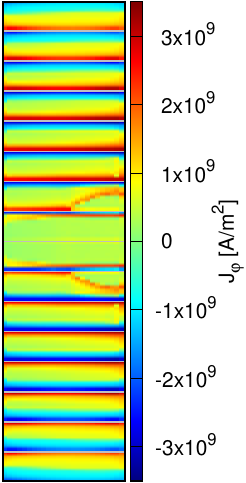}}\\
{\includegraphics[trim=148 8 150 9,clip,height=8.1 cm]{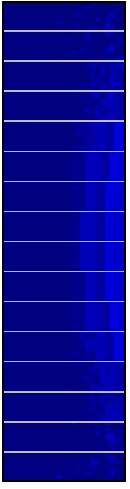}}
{\includegraphics[trim=148 8 150 9,clip,height=8.1 cm]{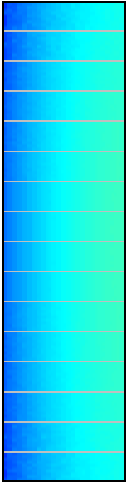}}
{\includegraphics[trim=148 8 150 9,clip,height=8.1 cm]{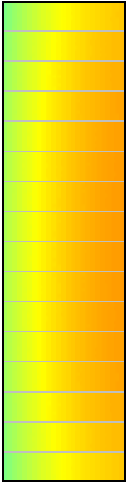}}
{\includegraphics[trim=148 8 102 9,clip,height=8.1 cm]{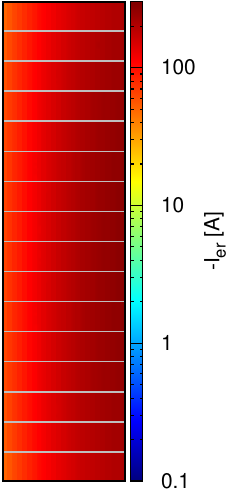}}
\caption{The same as figure \ref{f.Jph_ramp_666} but for the end of ramp down (time ``d'' in figure \ref{f.It}). Results for $I_m=666$ A and $\Rsur=10^{-6}$, $10^{-7}$, $10^{-8}$, $10^{-9}$ \Ohmmm from left to right.}
\label{f.Jph_down_666}
\end{figure}

\begin{figure}[tbp]
{\includegraphics[trim=148 11 150 9,clip,height=8 cm]{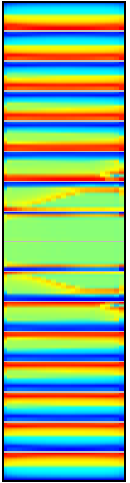}}
{\includegraphics[trim=148 11 150 9,clip,height=8 cm]{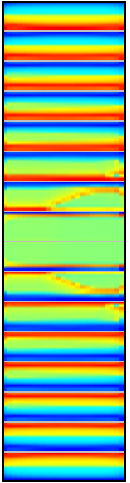}}
{\includegraphics[trim=148 11 150 9,clip,height=8 cm]{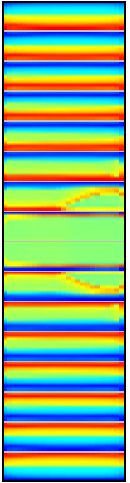}}
{\includegraphics[trim=148 11 94 9,clip,height=8 cm]{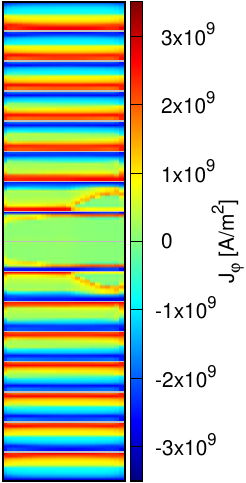}}
\caption{The same as figure \ref{f.Jph_ramp_666} but for the end of relaxation (time ``r'' in figure \ref{f.It}) and angular current density only ($I_m=666$ A and $\Rsur=10^{-6}$, $10^{-7}$, $10^{-8}$, $10^{-9}$ \Ohmmm from left to right).}
\label{f.Jph_end_666}
\end{figure}

The behavior of $J_\varphi(r,z)$ and $I_{er}(r,z)$ for the rest of the magnet charge and discharge process is essentially the same as for $I_m=466$~A but with much higher portion of turns with $I$ above their $I_c$ (see figures \ref{f.Jph_rel_666}-\ref{f.Jph_end_666}). The time evolution of $I_{r,{\rm av}}$ and $I_{\varphi,{\rm av}}$ is essentially the same as for $I_m=466$~A but with a much higher $I_{r,{\rm av}}$ at the plateau and faster convergence to  the static value of $I_{r,{\rm av}}$ (figure \ref{f.Iav}).

\subsection{Magnetic field at bore and screening current induced field}

\begin{figure}[tbp]
\centering
{\includegraphics[trim=48 0 50 5,clip,width=10 cm]{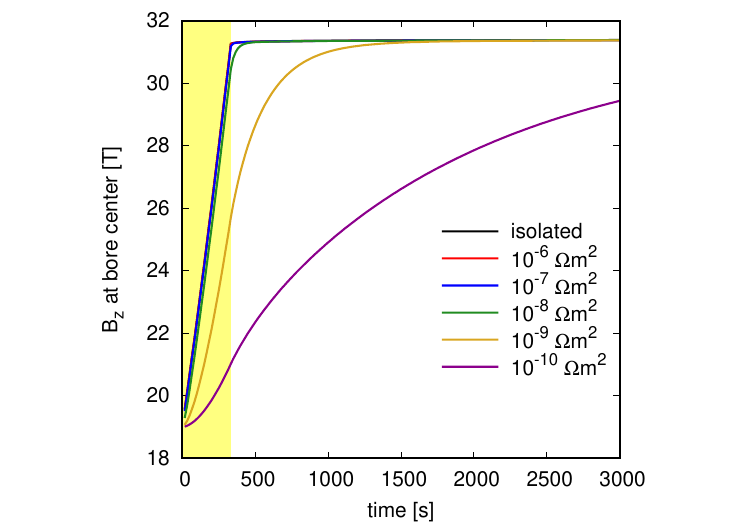}} 
\caption{Time evolution of the axial magnetic field, $B_z$, at the bore center for a maximum current of 333 A and several turn-to-turn \R{contact resistivities} ($10^{-6}$, $10^{-7}$, $10^{-8}$, $10^{-9}$, $10^{-10}$ \Ohmmm), as well as for the \R{insulated} coil configuration.}
\label{f.Bcen}
\end{figure}

For the considered ramp rate of 1 A/s and rated current ($I_m=333$~A), the generated magnetic field at the bore center for the metal-insulated coils ($\Rsur=10^{-6}$, $10^{-7}$~\Ohmmm) is practically the same as for the \R{insulated} coil (figure \ref{f.Bcen}), which is consistent with the average angular current in figure \ref{f.Iav}. Again, the non-insulated coil ($\Rsur=10^{-8}$~\Ohmmm) presents a delay in generated magnetic field, which becomes notorious with the soldered coil with $\Rsur=10^{-9}$~\Ohmmm (figure \ref{f.Bcen}). For all cases, the generated field converges to the \R{insulated} configuration after sufficiently long times after the initial ramp. In figure \ref{f.Bcen}, we also added the case of $\Rsur=10^{-10}$~\Ohmmm (very good soldering between turns), which requires plateaus of more than 1 hour to reach the stationary generated magnetic field. In any case, the stationary field is around 31.4 T, which is around 0.6 T below the design magnetic field of 32 T. The reason is the presence of screening currents, which generate Screening Current Induced Field (SCIF).

\begin{figure}[tbp]
\centering
{\includegraphics[trim=45 12 47 5,clip,width=10 cm]{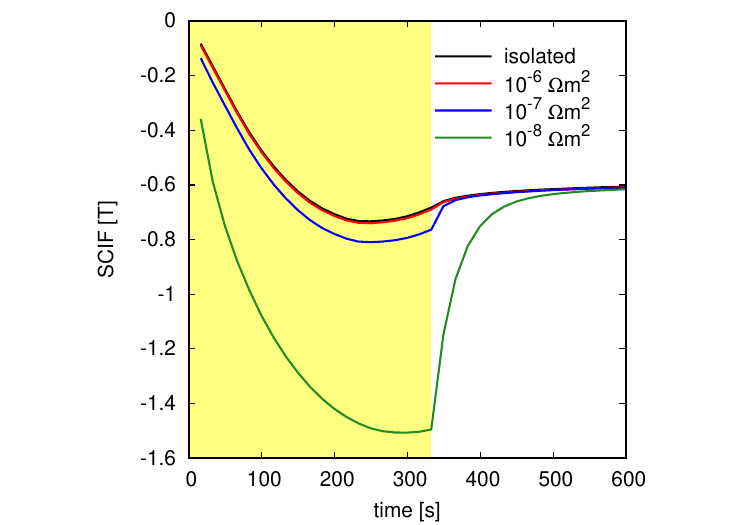}}\\
{\includegraphics[trim=45 0 47 5,clip,width=10 cm]{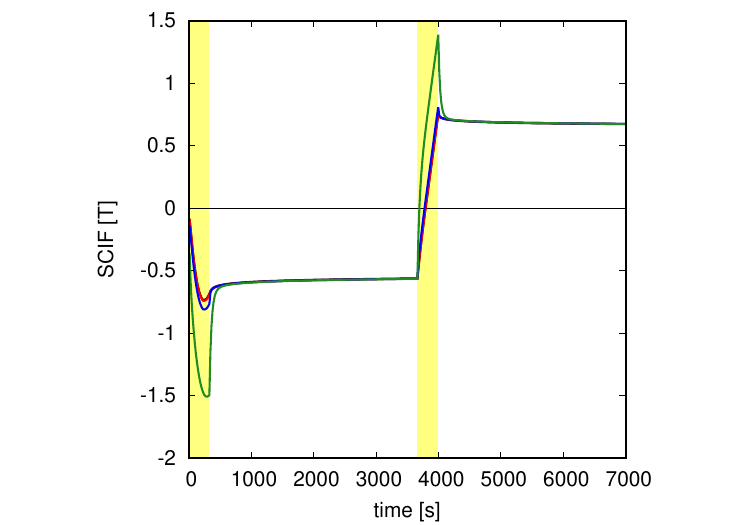}}
\caption{The screening current induced field (SCIF) at the bore center decreases with the turn-to-turn resistance ($10^{-6}$, $10^{-7}$, $10^{-8}$ \Ohmmm) and for $10^{-6}$ \Ohmmm  it is almost the same as for the \R{insulated} coil configuration. The top graph is a zoom close to the initial ramp of the graph in the bottom.}
\label{f.SCIF}
\end{figure}

In this article, we define the SCIF as the actual generated magnetic field minus the generated magnetic field for the \R{insulated} configuration with uniform $J_\varphi$. The SCIF for the metal-insulated coil with $\Rsur=10^{-6}$~\Ohmmm~is about the same as for the \R{insulated} coil (figure \ref{f.SCIF}), and hence this \R{contact resistivity} is an option if the magnet application requires small and stable SCIF. The case of $\Rsur=10^{-7}$ \Ohmmm~also presents relatively low increase in SCIF but the non-insulated coil, with $\Rsur=10^{-8}$~\Ohmmm, presents more than double SCIF, which is a high qualitative increase. The highest SCIF does not occur at the first ramp but after ramping down the HTS insert (figure \ref{f.SCIF}), where the static value is around 0.6 T. In figure \ref{f.SCIF}, we do not show the SCIF for the soldered coils, since this quantity is orders of magnitude larger than for the \R{insulated} case, and hence it does not make sense to consider the difference in generated magnetic field as SCIF. 

\subsection{AC loss}

\begin{figure}[tbp]
\centering
{\includegraphics[trim=49 0 50 5,clip,width=10 cm]{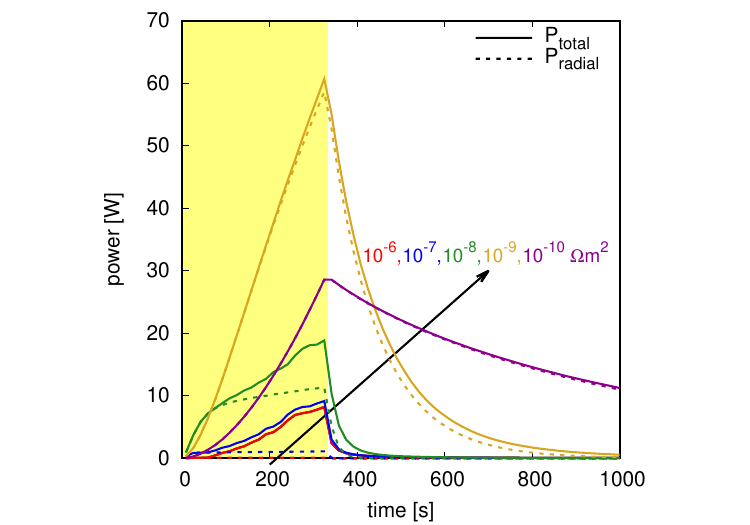}}
\caption{Power loss in the REBCO insert for the current time dependence of figure \ref{f.It} and $I_m=333$ A when considering several turn-to-turn \R{contact resistivities} in \Ohmmm . Solid lines are for the total power loss, while dash lines are for the component due to the radial currents only. \R{For $10^{-6}$ \Ohmmm, the dash line almost overlaps with the bottom graph border.}}
\label{f.P333}
\end{figure}

\begin{figure}[tbp]
\centering
{\includegraphics[trim=50 0 50 2,clip,width=10 cm]{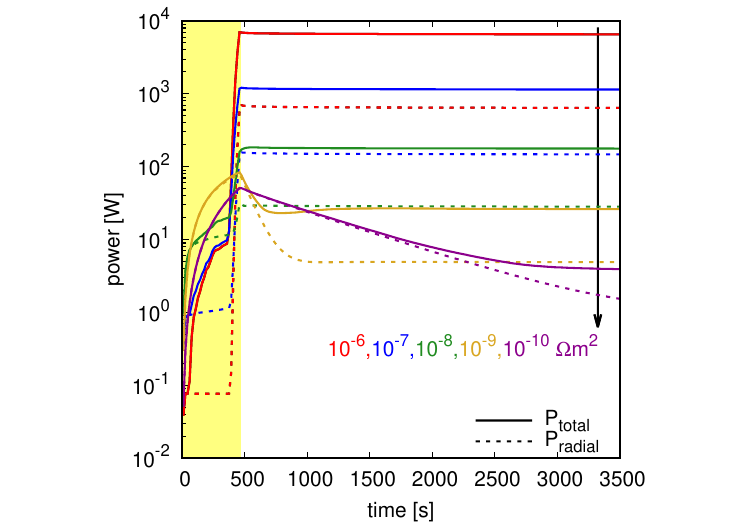}}
\caption{The same as figure \ref{f.P333} but for a maximum current of 466 A.}
\label{f.P466}
\end{figure}

\begin{figure}[tbp]
\centering
{\includegraphics[trim=50 0 50 2,clip,width=10 cm]{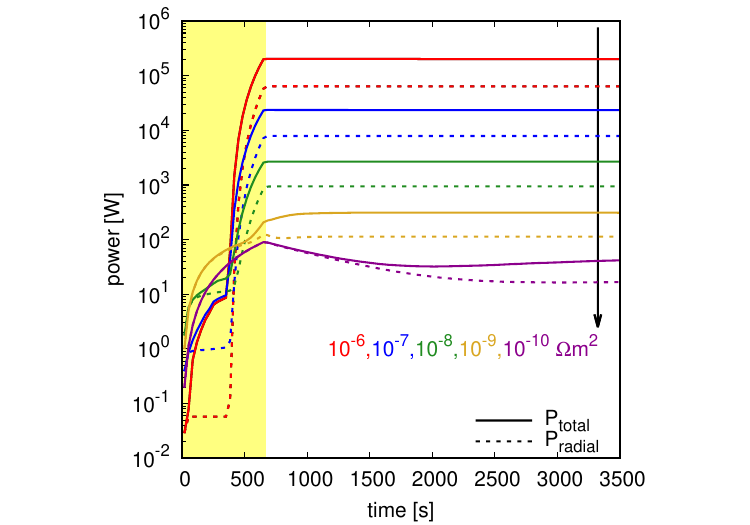}}
\caption{The same as figure \ref{f.P333} but for a maximum current of 666 A.}
\label{f.P666}
\end{figure}

The power AC loss can be separated between a contribution from the radial and angular currents, since the AC loss density follows $p = \vJ\cdot\vE = J_rE_r + J_\varphi E_\varphi$. Then, we define the ``radial loss'' as $P_{\rm radial}=\int_\Omega\dvol J_rE_r$ and the ``angular loss'' as $P_{\rm angular}=\int_\Omega\dvol J_\varphi E_\varphi$. For our case, the radial loss mainly occurs at the interface between turns, while the angular loss is mainly due to the superconducting tape. The total AC loss strongly depends on $\Rsur$ (figures \ref{f.P333}-\ref{f.P666}).

Now, we focus on the AC loss at operating current ($I_m=333$~A $=0.898 I_c$) (figure \ref{f.P333}). For metal-insulated coils ($\Rsur=10^{-6}$, $10^{-7}$~\Ohmmm) the radial loss is low or negligible, being most of the AC loss contribution from the superconducting screening currents. The non-insulated coil ($\Rsur=10^{-8}$~\Ohmmm) experiences around twice as much maximum power loss, being the radial loss higher than the angular loss. When decreasing further $\Rsur$ to $10^{-9}$~\Ohmmm, the AC loss highly increases, being almost entirely due to the radial currents. However, a further decrease of $\Rsur$ to $10^{-10}$~\Ohmmm decreases the maximum power loss, although the total dissipation energy ($Q=\int_0^\infty \dif tP(t)$) might be higher.

For a current above the coil $I_c$, $I_m=466$~A=1.26~$I_c$, the AC loss experiences a sharp rise for $I_m > I_c = 371$~A for the metal-insulated and non-insulated coils ($\Rsur=10^{-6}$, $10^{-7}$, $10^{-8}$~\Ohmmm) (figure \ref{f.P466}). This is due to the sharp increase in the superconductor loss, which impacts to the angular loss. Indeed, for $I$ beyond $I_c$, the angular loss is dominant, although the radial loss is substantially large (notice the logarithmic scale). In this range of $\Rsur$, the total loss increases with $\Rsur$ because the transfer from angular to the radial current decreases. Interestingly, the angular loss roughly increases proportionally to $\Rsur$, experiencing huge power loss at $\Rsur=10^{-6}$~\Ohmmm. The largest power loss, of 7000 W, corresponds to a resistive voltage of the power source, $V_{\rm res}=P/I$, of around 15 V. This voltage is relatively large but feasible for certain power sources for magnets. However, either intended or accidental voltage limitation of the source could cap the power loss. 

The qualitative behavior for the soldered coils ($\Rsur=10^{-9}$, $10^{-10}$~\Ohmmm) is very different. At the initial ramp, the angular radial loss dominates also for $I_m> I_c$ (figure \ref{f.P466}). The reason is that $I_\varphi$ does not reach $I_c=371$ A (see figure \ref{f.Iav}). The stationary angular current is not reached until until relatively long times after the ramp. When $I_\varphi>I_c$ there is a mild rise of the total loss and the angular loss becomes dominant. A remarkable feature of soldered coils is that there is not a high increase of the AC loss compared to the under-critical case (compare figures \ref{f.P333} and \ref{f.P466}). In addition, both the peak power and dissipation energy increase with $\Rsur$. As well, the stationary loss is roughly proportional to $\Rsur$.

The case of $I_m$ well above $I_c$ ($I_m=666$ A$=1.80I_c$) is extreme for metal-insulated coils (figure \ref{f.P666}). Indeed, for $\Rsur=10^{-6}$ \Ohmmm the computed power loss is around 200 and 20 kW for $\Rsur=10^{-6}$ and $10^{-7}$~\Ohmmm. This corresponds to a source resistivity voltage of $V_{\rm res}$ of around 300 and 30 V, which is very high for magnet power sources. Then, for this current source voltage limitation will play a crucial role for metal-insulated coils. The qualitative behavior of non-insulated coils ($\Rsur=10^{-8}$ \Ohmmm) is the same but with much lower power loss for $I$ well above $I_c$. For all cases, the stationary loss at the end of the plateau is roughly proportional to $\Rsur$, and hence it varies by orders of magnitude. For $\Rsur=10^{-9}$ \Ohmmm, there is a moderate increase in AC loss with $I$ for $I>500$ A but it is much less pronounced than the metal-insulated and non-insulated coils. Finally, the case of very good soldering between turns, $\Rsur=10^{-10}$ \Ohmmm presents the lowest AC loss and dissipation energy. The fact that both the maximum and stationary loss are moderate, 649 and 42.6 W respectively, might enable operation at these large currents	. In that case, the generated magnetic field at the plateau could increase from 31.4 T to 36.7 and 42.6 T, for $I=466$ and $666$ A, respectively. Naturally, other factors could limit the generated magnetic field, such as self-heating or high mechanical stress.

\section{Conclusion}

In this article, we have presented a fast and accurate numerical method in order to compute screening currents and radial currents in metal-insulated, non-insulated and soldered REBCO magnets. \R{Our computations are really fast, since the computing time for the evolution from $I$=0 to the end of the initial ramp is around 56 and 200 s for 10 and 20 elements per homogenized turn in a standard table computer or laptop. That is less than the real operation time of the magnet, being 333 s for a relatively high ramp of 1 A/s.} The method is based on an axi-symmetric approximation, assuming that both the angular and radial current densities, $J_\varphi$ and $J_r$, are independent on the angular coordinate. We applied this principle to the Minimum Electro-Magnetic Entropy Production (MEMEP) \cite{pardoE2015SST, pardoE2017JCP} and benchmarked the results with a different method based on $A-V$ formulation and programmed in Matlab, showing very good agreement. 

We also systematically analyzed a REBCO insert for a 32 T magnet design. In particular, we studied $J_r$, $J_\varphi$, the generated magnetic field, the Screening Current Induced Field (SCIF), and the AC loss. We have shown that metal-insulated coils enable transfer of angular current in the radial direction, and hence magnet protection, while keeping the same screening currents and SCIF of \R{insulated} coils, even at relatively high ramp rates of 1 A/s. \R{For currents below the critical current of any turn of the insert, the AC loss decreases with the turn-to-turn resistance until it reaches the same level as \R{insulated} coils. If the current overcomes the critical current of the turns for a significant portion of the insert, both metal-insulated and insulated coils present high AC loss, as expected.} Surprisingly, soldered coils with low resistance between turns present relatively low AC loss for over-current configuration, which might enable higher generated magnetic fields. However, the stable generated magnetic field is reached after very long times after the ramp (or for very slow ramps). This strongly limits the rate in soldered coils.

The numerical method presented here can be applied to optimize high-field magnets regarding SCIF and required relaxation times in metal-insulated or non-insulated coils. It also serves as the basis for future electro-thermal modelling and multi-physics modeling that also includes mechanical properties.

\section{Acknowledgements}

We acknowledge Oxford Instruments for providing details on the cross-section of the LTS outsert. This project has received funding from the European Union's Horizon 2020 research and innovation programme under grant agreement No 951714 (superEMFL), and the Slovak Republic from projects APVV-19-0536 and VEGA 2/0098/24. Any dissemination of results reflects only the authors' view and the European Commission is not responsible for any use that may be made of the information it contains.

\end{document}